\documentstyle[prb,aps,multicol,epsf,graphics]{revtex}

\begin{document}
\draft
\newcommand {\nn}{\nonumber}
\newcommand{\boxed}[1]{\fbox{$\displaystyle{#1}$}}
\newcommand{\Hi}{\mbox{${\cal H}$\/}}
\newcommand{\Cr}{\mbox{$A^*$}}
\newcommand{\Spur}{\mbox{Spur}}
\newcommand{\bra}[1]{\mbox{$ \langle #1 | $}}
\newcommand{\ket}[1]{\mbox{$ | #1 \rangle $}}
\newcommand{\braket}[2]{\mbox{$ \langle #1 | #2 \rangle $}}
\newcommand{\expect}[1]{\mbox{$ \langle #1 \rangle $}}
\newcommand{\mat}[3]{\mbox{$ \langle #1 | #2 | #3 \rangle $}}
\newcommand{\An}{\mbox{$A$}}
\newcommand{\Va}{\mbox{$\Omega$\/}}
\newcommand{\Real}{\mbox{$I{\!}R$\/}}
\newcommand{\Complex}{\mbox{$/{\!\!}C$\/}}
\newcommand{\Poincare}{\mbox{$P_+^\uparrow$\/}}
\newcommand{\Wi}{\mbox{${\cal W}$\/}}
\newcommand{\Schwartz}{\mbox{${\cal S}\/$}}
\newcommand{\Fock}{\mbox{${\cal F}\/$}}
\newcommand{\srcacuo}
        {\mbox{$\mbox{Sr}_{14-x}\mbox{Ca}_{x}\mbox{Cu}_{24}\mbox{O}_{41}$}} 
\newcommand{\srtcacuo}
        {\mbox{$\mbox{Sr}_{3}\mbox{Ca}_{11}\mbox{Cu}_{24}\mbox{O}_{41}$}} 
\newcommand
{\sracuo}{\mbox{$\mbox{Sr}_{14-x}\mbox{A}_{x}\mbox{Cu}_{24}\mbox{O}_{41}$,
        (A=Ca, Y)}}
\newcommand{\cuoplane}{\mbox{$\mbox{Cu}_{2}\mbox{O}_{3}$}}
\newcommand{\srcuo}{\mbox{$\mbox{Sr}\mbox{Cu}_{2}\mbox{O}_{3}$}}
\newcommand{\ad}{a^{\dag}}
\newcommand{\bd}{b^{\dag}}
\newcommand{\bcd}{\tilde{c}^{\dag}}
\newcommand{\bc}{\tilde{c}}
\newcommand{\ba}{{\mathbf a}}
\newcommand{\bi}{{\mathbf i}}
\newcommand{\bj}{{\mathbf j}}
\newcommand{\bk}{{\mathbf k}}
\newcommand{\bnot}{{\mathbf 0}}
\newcommand{\bsigma}{\frac{\mbox{\boldmath$\sigma$\unboldmath}}{2}}
\newcommand{\bsigmad}{\frac{\mbox{\boldmath$\tilde{\sigma}$\unboldmath}}{2}}
\newcommand{\bsigmadm}{\frac{\mbox{\boldmath$\tilde{\sigma}$\unboldmath}_{\mu}}
{2}}
\newcommand{\bun}{\mbox{\boldmath$1$\unboldmath}}
\newcommand{\bS}{{\mathbf S}}
\newcommand{\kb}[1]{b_{#1}^{\dag}}
\newcommand{\bb}[1]{b_{#1}}
\newcommand{\bv}{{\mathbf v}}
\newcommand{\cd}{c^{\dag}}
\newcommand{\cac}[2]{\cd_{#1}\ba c_{#2}}
\newcommand{\caccac}[4]{\cac{#1}{#2}\cac{#3}{#4}}
\newcommand{\hc}{{\mathrm h.c.}}
\newcommand{\Jperp}{J_{\perp}}
\newcommand{\tperp}{t_{\perp}}
\newcommand{\Jpar}{J_{\|}}
\newcommand{\cNk}{{\cal N}}
\newcommand{\cSk}{{\cal M}}
\newcommand{\tpar}{t_{\|}}
\newcommand{\SSe}[2]{\expect{\hat{S}^{\dag}(#1)\hat{S}(#2)}}
\newcommand{\TTo}[2]{\expect{\hat{T}_{0}^{\dag}(#1)\hat{T}_{0}(#2)}}
\newcommand{\TTu}[2]{\expect{\hat{T}_{\uparrow}^{\dag}(#1)\hat{T}_{\uparrow}(#2)}}

\title{Phase Diagram of Coupled Ladders}

\author{ T.~F.~A.~M\"uller and T.~M.~Rice}

\address{Institut f\"ur Theoretische Physik, \\
	ETH-H\"onggerberg, CH-8093 Zurich, Switzerland}
\date{\today}
\maketitle
\begin{abstract}
The 2-leg \mbox{$t$--$J$} ladder forms a spin liquid at half-filling which evolves to
a Luther-Emery liquid upon doping. Our aim is to obtain a complete phase
diagram for isotropic coupling (i.e. rungs and legs equal) as a function of
electron density $n$ and the ratio $J/t$. Two known limiting cases are:
\mbox{$n<1/2$}
which is a single band Luttinger liquid and small hole
doping \mbox{$\delta \ll 1$} for \mbox{$J/t
\longrightarrow 0$} which is a Nagaoka 
ferromagnet. Using Lanczos techniques we examine the region between the Nagaoka
and Luther--Emery phases for \mbox{$1>n>1/2$}. We find evidences for gapless behavior
in both spin and charge channels for \mbox{$J/t<0.3$} 
consistent with Luttinger liquids in both bonding and
anti-bonding bands (i.e., C2S2). This proposal is based on the behavior of
spin and charge correlation functions. For example the hole--hole correlation
function which displays hole pairing at larger $J/t$, shows hole--hole
repulsion in this region. As a further test, we examined the dependence of the
energy on a relative phase shift between bonding and antibonding bands. For 
$J/t < 0.3$ this is very weak, indicating a lack of pairing between 
these
channels. 
\end{abstract}

\preprint{ETH-TH/98-09}

\begin{multicols}{2}
\narrowtext
\section{Introduction}
	The surprising discovery of high-$T_c$ materials by  Bednorz and
M\"uller 
\cite{bednorz} has sparked renewed interest 
in low-dimensional strongly correlated quantum systems. A striking feature 
of these materials is that they show simple long range
antiferromagnetic (AF) order at low temperatures when they are not doped with
holes. 

Between the  well-known one-dimensional systems and the difficult two
dimensional system, coupled chains (or ladders) are
interesting intermediate systems. The most striking feature of 2m-leg
ladders is the appearance of a spin-gap at half-filling and small doping
\cite{dagotto,barnes,whiteno,gopa,Troyer,tsunetsugu,frishmuth}. 
	For instance compounds such as Sr$_{n-1}$Cu$_{n+1}$O$_{2n}$ 
have been shown 
to be well described by a lattice of coupled \mbox{$\nu_n$}-leg ladders,
with \mbox{$\nu_n=(n+1)/2$} \cite{ricegosi,dagottorice}.
Another example containing two-leg ladders
is the system \srcacuo \cite{carron,siegrist} which is a 
material with doped ladders \cite{osafune,mizuno}
and where superconductivity (under high pressure)
has been observed\cite{magishi,uehara,Motoyama}. 

	An important step is the
determination
of the phase diagram. 
Balents and Fisher \cite{BalentsFisher} have computed the
phase diagram 
of the Hubbard-ladder in the weak coupling limit \mbox{$U \longrightarrow 0$} in the
frame of bosonization and renormalization group theory. 
To distinguish the different phases they  introduced 
the notation C$x$S$y$ for labeling phases with $x$ {\it gapless} charge
and $y$ {\it gapless} spin excitations. 
	Noack {\it et al.} \cite{Noacketal} investigated then the different
phases numerically with Density Matrix Renormalization Group (DMRG) methods for
 intermediate coupling
and found good agreement with the analytical
results of the weak coupling limit. 
The main feature is the existence of two regions, the Luther-Emery (LE)
region with a spin 
gap (C1S0) and the Luttinger-Liquid (LL) with one gapless excitation (C1S1).

	The strong coupling \mbox{$t$--$J$} model \cite{anderson,ZhangRice} has also
been studied extensively by different  
authors. It is given by the Hamiltonian
\begin{eqnarray}
	 H & = & 
	  \tperp \sum_{j\sigma} \left( \bcd_{1j,\sigma} \bc_{2j,\sigma} +
\hc  \right) \\ \nonumber & &
	 +\tpar \sum_{lj\sigma} \left( \bcd_{lj+1,\sigma} \bc_{lj,\sigma} +
\hc \right) \nonumber   \\
	 & & +\Jperp \sum_{j} \left( \bS_{1j} \bS_{2j}
-\frac{1}{4}n_{1j}n_{2j} \right) \nonumber 
	\\  & &
	  +\Jpar \sum_{lj} \left( \bS_{lj+1} \bS_{lj}
-\frac{1}{4}n_{lj+1}n_{lj} \right),
\label{hamiltonian} 
\end{eqnarray}
where the index $l \in \{1,2 \}$ refers to legs and $j$ to rungs. 
The $\bc$ operators denote the fermion operators with
projection onto the singly occupied states, i.e.
\mbox{$\bc_{lj,\sigma}=c_{lj,\sigma}(1-n_{lj,-\sigma})$}. 
In this paper, we report a detailed
study of the phase diagram for isotropic coupling (i.e. rungs and legs equal) as a function 
of electron density $n$ and the ratio $J/t$.

In the strong coupling limit \mbox{$\Jperp \gg \Jpar$}, the spin-gap can be easily represented.
At half-filling,  the ground state is formed by spin-singlets lying on each rung. 
Turning over one spin gives a triplet on the corresponding rung.
The energy difference between the two states, the spin-gap, is
$\Delta \simeq \Jperp$.
Numerically, it is found that the spin gap in the isotropic case
$\Jpar=\Jperp$ reduces to \mbox{$\Delta \approx \Jperp/2$}
\cite{dagotto,barnes,whiteno}. 
The spin-gap of a doped ladder for $\Jperp \gg \Jpar$ is due to a qualitatively different process.
Some singlets have been replaced by hole-pairs moving along
the ladder with a renormalized hopping  $\widetilde{\tpar}$. The lowest excitation arises 
by breaking one hole pair (or equivalently a singlet bond) 
and putting each unbound spin on well-separated
rungs. Turning over one of the separate spins gives the lowest triplet excitation with
$\Delta \approx \Jperp -2\tperp -2\tpar$  leading to a finite spin-gap 
(LE). 
At the isotropic point $\Jperp=\Jpar$ and for $J/t \geq 0.3 $ the spin-gap
still exists where
 now the bound holes share neighboring rungs and
 strong singlet correlations are measured on the remaining
sites~\cite{Troyer,WhiteScal}. 
The spin-gap is thus a discontinuous function of doping in the vicinity of
half-filling.
	For higher values of the parameter $J/t$,
holes and electrons separate completely at all doping levels.

Thus, the appearance of the spin gap in a doped ladder is directly correlated
with the formation of hole pairs, and it is an interesting problem to
study their 
stability. Poilblanc {\it et al.} \cite{PoilblancScal} used a numerical method
based on Lanczos  
algorithms, which has allowed them to distinguish between the gapped and
gapless
regions in the phase diagram. Hayward and Poilblanc \cite{HaywardPoil}
 determined the non-universal correlation
exponents \mbox{$K_{\rho}$} defining the behavior of the long-range correlations. 
They found at low electron density \mbox{$n<1/2$} the system is
in a LL phase, while for a higher electron density a
gapped phase with hole pairs is stabilized. 
A large region of this
gapped phase exhibits dominant superconducting correlations. 
 The boundary of these two phases was determined to be \mbox{$n \simeq 1/4$}
 where, in the band picture, the Fermi energy $E_F$ just touches
the antibonding band.

	However, some parts of the phase diagram are still unclear. First,
the
finite-size scaling process to determine the spin-gap region 
does not give clear results for values of \mbox{$J/t<0.3$}. 
Second, the radius of the hole pairs increases with decreasing $J/t$,
and it is not clear whether it diverges for a particular value, and if a
transition to a gapless phase occurs for small $J/t$. 
For very small values of $J/t$, the $t$--$J$ model is 
similar to the Hubbard model with \mbox{$U\rightarrow \infty$}. 
Since for two coupled chains, 
all spin configurations with a fixed $S_z$ are
coupled by hole hopping as in a two-dimensional system, the essential
condition for 
the Nagaoka theorem \cite{Nagaoka} is fulfilled.
A ferromagnetic phase occurs
for very small values of $J/t$ and low hole doping. 
In this phase, no spin gap occurs and holes
repel each other. With increasing $J/t$, 
the ground state  rapidly evolves into a singlet state. 

This part of the phase diagram will be extensively studied on the basis
of exact diagonalization results for small clusters, typically 10-rung
ladders. Finite-size effects can be important for such systems, and 
it will be tried to minimize them as much as possible. 

\section{Boundary Conditions}
\label{bouncon}
	A Lanczos algorithm will be used to investigate the different
phases. With current computers it is possible to investigate 2 leg ladders
of length $L=10$ at any filling. 
In order to carry out a systematic analysis
for different doping, the boundary conditions (BC) must be chosen
carefully. 
 
\subsection{Definition of Boundary Conditions}
	Usually, the BC are defined 
 in the non-interacting limit of the model 
where the Hamiltonian is exactly described by two parallel
bands \mbox{$E_{\pm}(k)$} also labeled  \mbox{$E(k_x,k_y)$} with 
$k_y \in \{ 0, \pi \}$.
	Since the system is finite, the \mbox{$\bk(=(k_x,k_y))$-values} 
belong to  a discrete set.
In general $k_x=\frac{2\pi}{L}l+\phi$, where {\it a priori} 
$\phi$ needs not be the same for bonding ($k_y=0$)  and antibonding ($k_y=\pi$)
branches.
BC giving either fully
occupied or empty single particle orbitals in the non-interacting case,
called closed shell  (CS) boundary conditions (CSBC), are chosen.

	If this condition is not fulfilled, the ground state is degenerate
This favors the pairing of spins at $E_F$ and enhances
pairing instabilities. This configuration is called
open shell boundary conditions (OSBC).

\begin{figure}
\centerline{\epsfysize=1.5cm\epsfbox{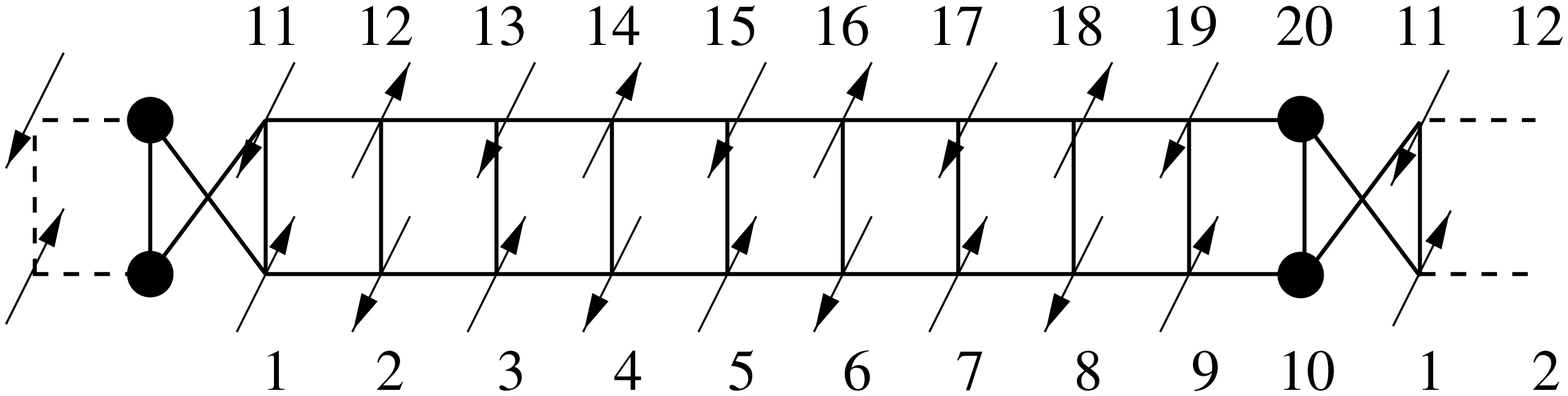}}
\vspace{5mm}
\caption{A 10-rung ladder with two holes on the same rung. Due to MBC
The hole pair when sitting on the same rung do not frustrate the N\'eel
ordering. The labeling
convention used in future graphs is shown.  }
\label{mbc10h2}
\end{figure}

The definition of the BC for the bonding and antibonding operators
\begin{eqnarray}
 b_{j\sigma} & = & \frac{1}{\sqrt{2}}(c_{1j,\sigma}+c_{2j,\sigma}), \\
 a_{j\sigma} & = & \frac{1}{\sqrt{2}}(c_{1j,\sigma}-c_{2j,\sigma}),
\end{eqnarray}
are given by the set of equations (the spin index is droped for simplicity)
\begin{eqnarray}
	T b^{\dag}_{j} T^{-1} & = & e^{i\phi} b^{\dag}_{j+1},\ \ \ 1 \le j <
L,\\ 
	T a^{\dag}_{j} T^{-1} & = & e^{i\phi} a^{\dag}_{j+1},\  \ \ 1 \le j <
L, \\
	T b^{\dag}_{L} T^{-1} & = & e^{i\phi} b^{\dag}_{1}, \\
	T a^{\dag}_{L} T^{-1} & = & e^{i(\phi+m\pi)} a^{\dag}_{1}. 
\end{eqnarray}
For $m=0$ the usual BC with the geometry of a ring are recovered. They 
will be generally referred
to as RBC($\phi L$). \mbox{$\phi L=0(\pi)$} are the most-used phases, which will be
called periodic(antiperiodic) boundary condition or PBC(APBC).
For $m=1$ the 
operators at the end of the legs fulfill the relationship
\begin{eqnarray}
	T c^{\dag}_{1\, L} T^{-1} & = & e^{i\phi} c^{\dag}_{2\, 1}, \\
	T c^{\dag}_{2\, L} T^{-1} & = & e^{i\phi} c^{\dag}_{1\, 1},
\end{eqnarray} 
corresponding to the geometry of a Moebius band. They
will be called Moebius Boundary Conditions and denoted by MBC$(\phi L)$.
MBC($0$) means that periodic
boundary conditions for bonding states and anti-periodic boundary conditions 
for antibonding states are taken and vice-versa for MBC($\pi$).  In real
space they can be viewed as a way to prevent antiferromagnetic frustration in
some cases as schematically shown in Fig.~\ref{mbc10h2} for 
a 10-rung ladder with two holes.
For a phase differences $m\pi$ other than $0$ or $\pi$ the
translation of one particle leads to two states, each one with one particle on
each leg. This case will not be considered further.

By introducing these BC it is 
possible to get CSBC for all doping with an even number of holes. In 
the situation where two CSBC are possible, 
the one giving the minimal energy {\it in the \mbox{$t$--$J$} model} will
be generally preferred \cite{lowestBC}.

\begin{figure}
\centerline{\epsfysize=7.cm\epsfbox{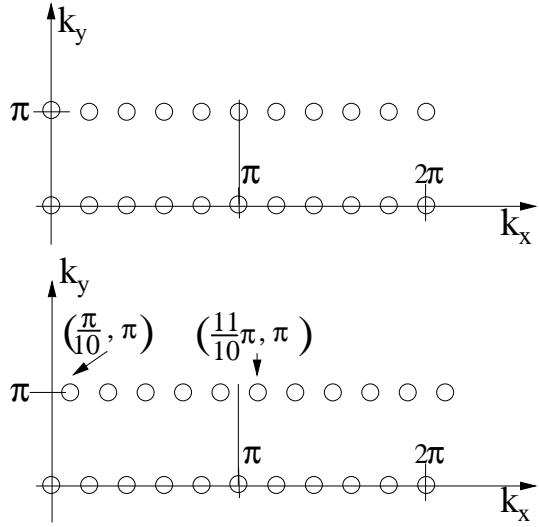}}
\vspace{5mm}
\caption{The two different sets of $k$ values for RBC and MBC for a 10-rung
ladder. 
For MBC, the set of $k$ values at $k_y=\pi$ is shifted by 
$\pi/L$ respectively to the $k_y=0$ branch.}
\label{Fourier}
\end{figure}

The Fourier transform
\begin{equation}
	f_k=\sum_{\bj} e^{i\bk\bj}f_{\bj},
\end{equation}
with \mbox{$\bk=(k_x,k_y)$} and \mbox{$\bj=(j_1,j_2)$} will be computed 
consistently 
by taking $k_x$ as a function of \mbox{$k_y\in \{ 0, \pi\}$},
\begin{equation}	
	k_x(k_y)=\left\{ \begin{array}{ll}
		\frac{2\pi}{L}l &  \mbox{RBC},\\[2mm]
		\frac{2\pi}{L}l+\frac{\pi}{L}\delta_{k_y,\pi} & \mbox{MBC},
			 \end{array}
			\right.
\end{equation}
where $l$ is an integer. The two 
 possible sets of $\bk$-values are summarized
in  Fig.~\ref{Fourier}.

\section{Nagaoka Phase}
\label{snag}
As discussed in the introduction, a Nagaoka phase~\cite{Nagaoka}
for $J/t\longrightarrow 0$ is expected at low doping.
	Numerically, Troyer \cite{Troyer} has shown that 
the ground state of  $L\ge 4$-rungs \mbox{$t$--$J$} ladders with a doping of
two holes is a saturated ferromagnet at $J/t=0.$  
In the following, the doping and $J/t$ dependence of this phase will
be investigated. 

\begin{figure}
\centerline{\epsfxsize=8.5cm\epsfbox{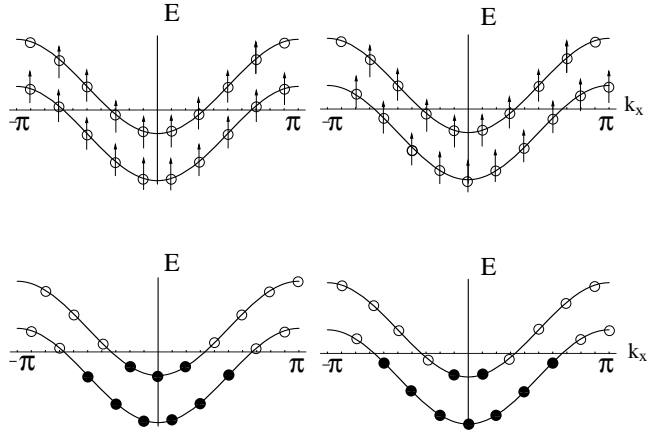}}
\vspace{1mm}
\caption{The CSBC for a 10-rung ladder with two holes. 
The upper graphs represents the filling of the band with 
APBC and MBC$(0)$ for a saturated ferromagnet. 
The lower graphs shows the filling in the
non-interacting limit for $S=0$ for MBC$(\pi)$ and MBC$(0$). 
Filled circles represent
the doubly occupied states while  white circles stand for empty
states.}
\label{Closedshell}
\end{figure}

	The analysis of that phase is simplified by first considering
the subspace of the completely polarized states (\mbox{$S_z^{tot}=N_s/2$} with
$N_s$ the number of spins)
which is equivalent to a spinless fermion system. The 
eigenvalues are exactly given by the band picture $E_{\pm}(\bk)=-2\tpar\cos(k_x)
\pm \tperp$ and the eigenstates are direct products of bonding and
antibonding states. For finite clusters CSBC  
can be sometimes obtained with different BC. For instance, 
for a 10-rung ladder with 2 holes and with \mbox{$\tperp=\tpar$} both
APBC and MBC($0$) allow a CS configuration with the same
ground state energy as schematically plotted 
in the upper graphs of Fig.~\ref{Closedshell}.
For higher \mbox{$J/t$}, when the ground state is a singlet, 
the corresponding CSBC configurations
are shown in the lower graphs of Fig.~\ref{Closedshell} where black circle
stand for fully occupied states. Both
MBC($0$) and  MBC($\pi$) are possible.

\begin{figure}
\centerline{\rotatebox{-90}{\epsfysize=9.cm\epsfbox{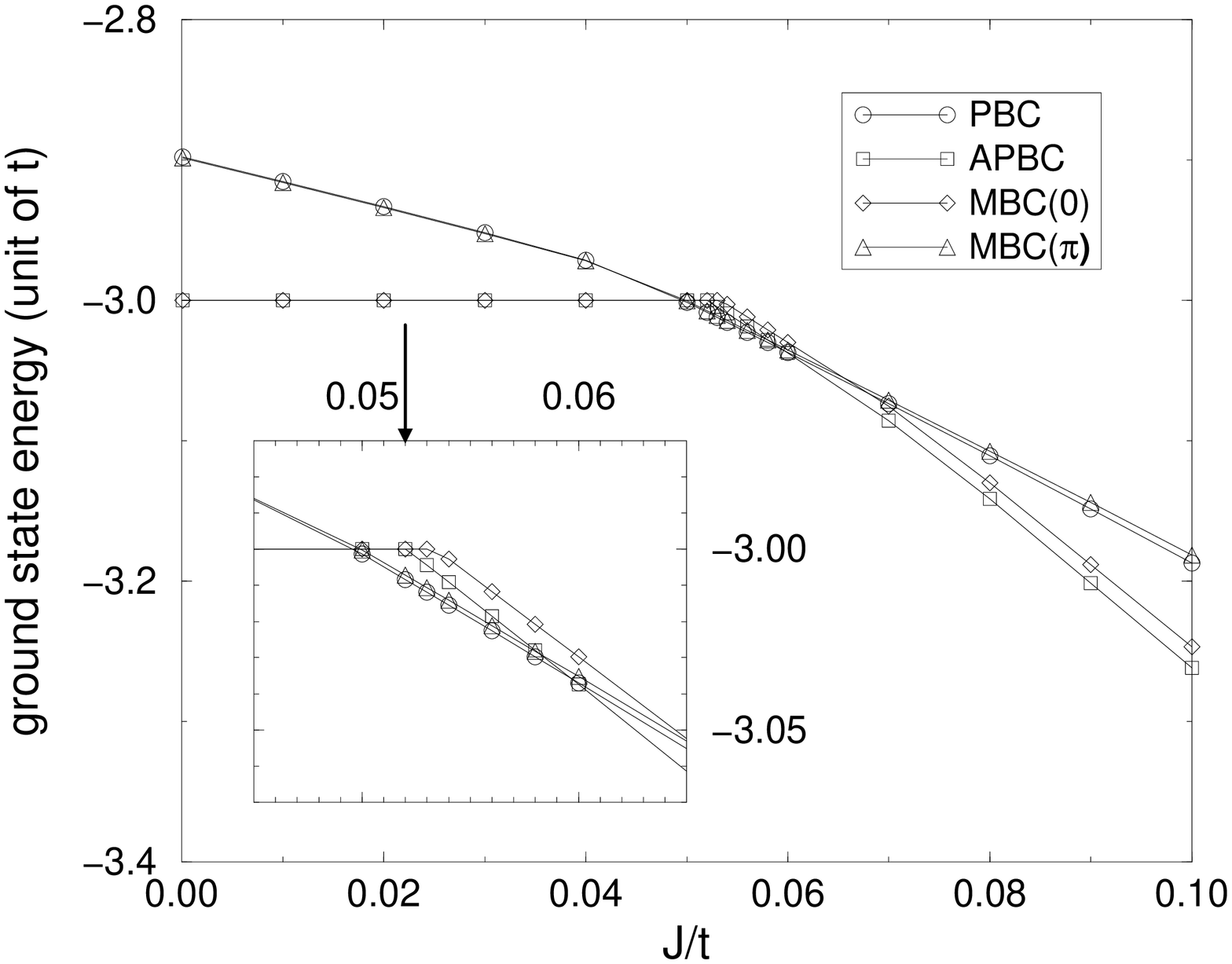}}}
\centerline{\rotatebox{-90}{\epsfysize=9.cm\epsfbox{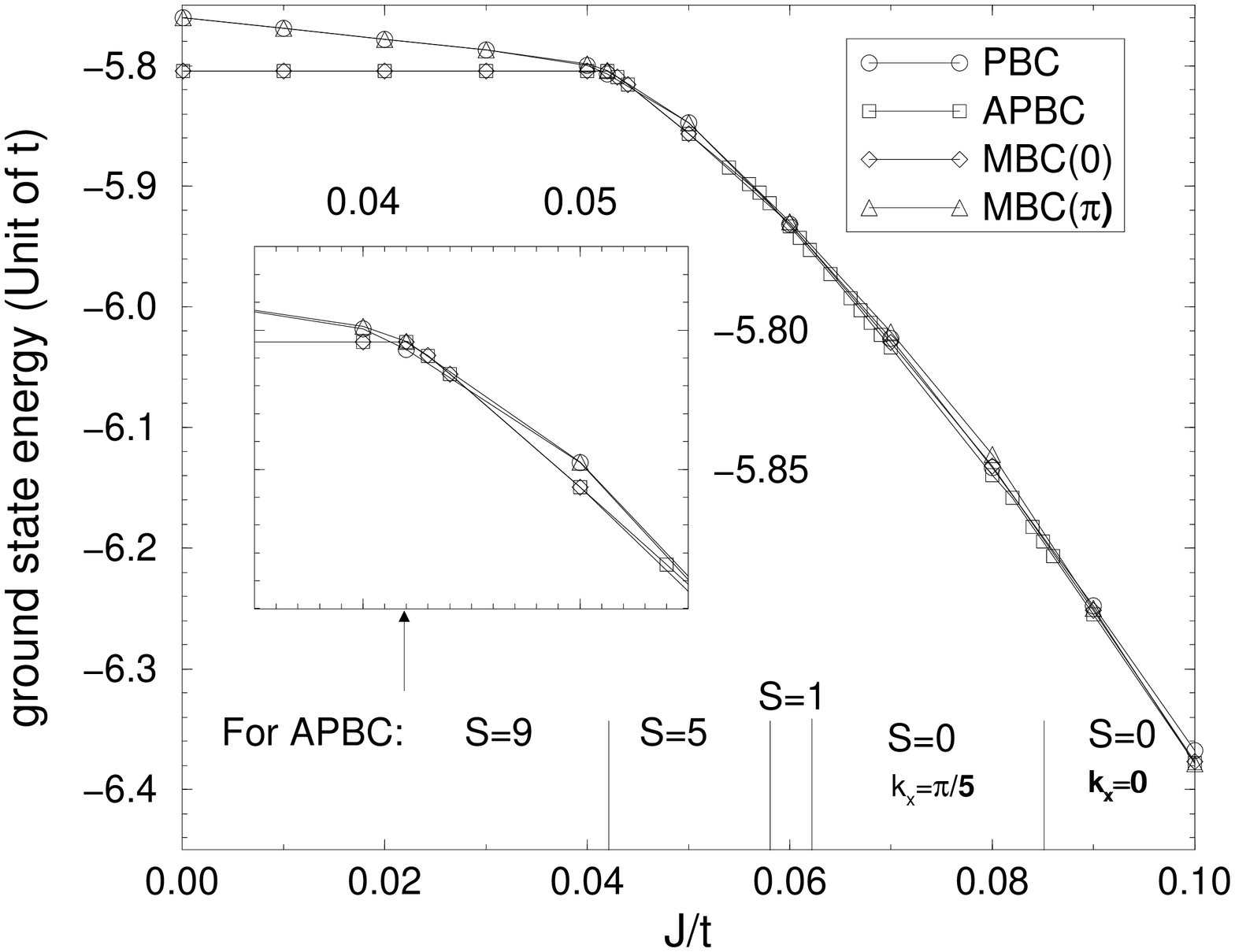}}}
\vspace{1mm}
\caption{The upper(lower) 
graph shows the data for a 5(10)-rung ladder with 1(2) hole(s)
for different BC. The insets show
the magnified region where the saturated ferromagnet appears. The
total spin of the 
ground state (using APBC) for the 10-rung ladder is indicated  
at the bottom of the lower graphs.}
\label{ferrened1} 
\end{figure}

\begin{figure}
\centerline{\rotatebox{-90}{\epsfysize=9.cm\epsfbox{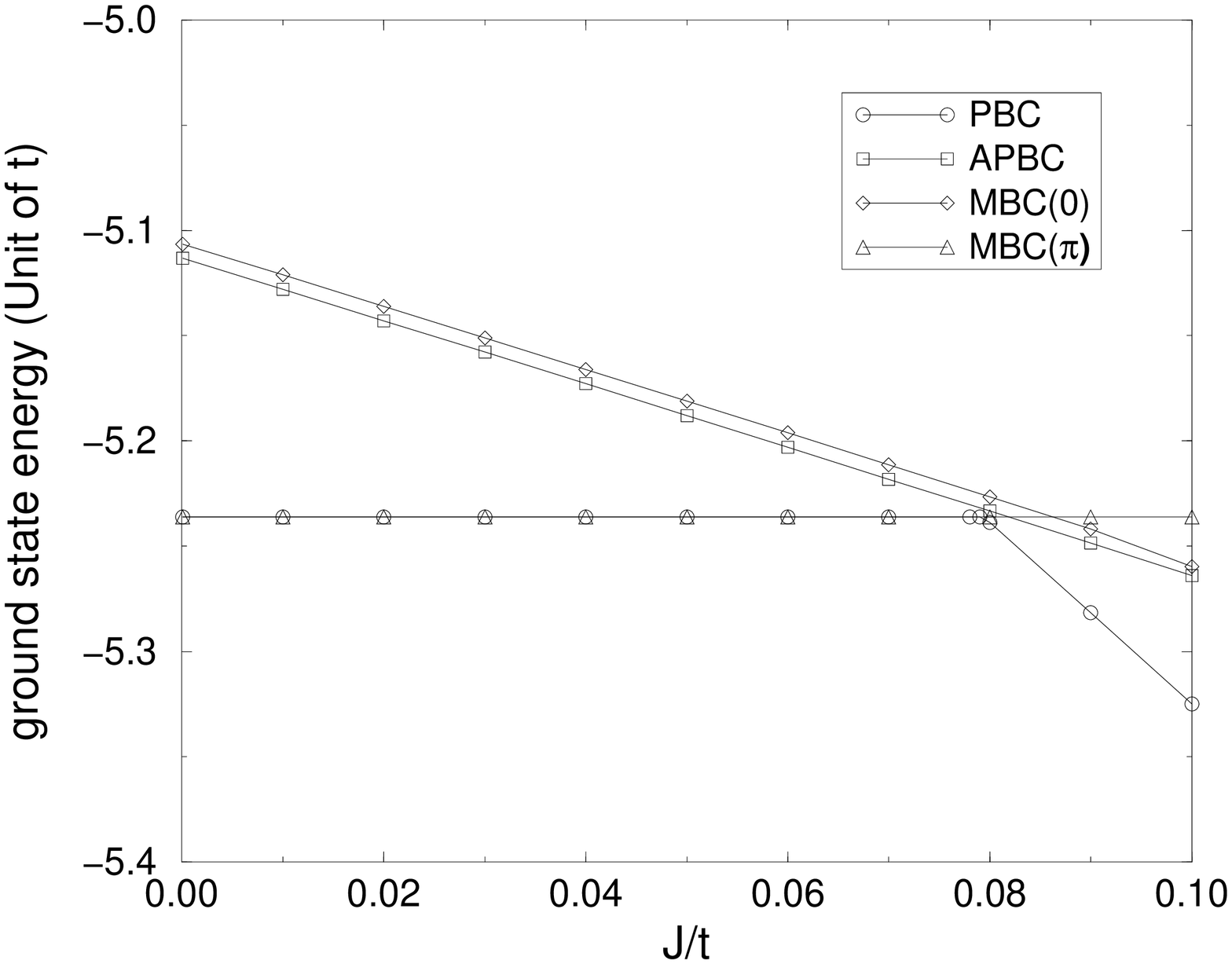}}}
\centerline{\rotatebox{-90}{\epsfysize=9.cm\epsfbox{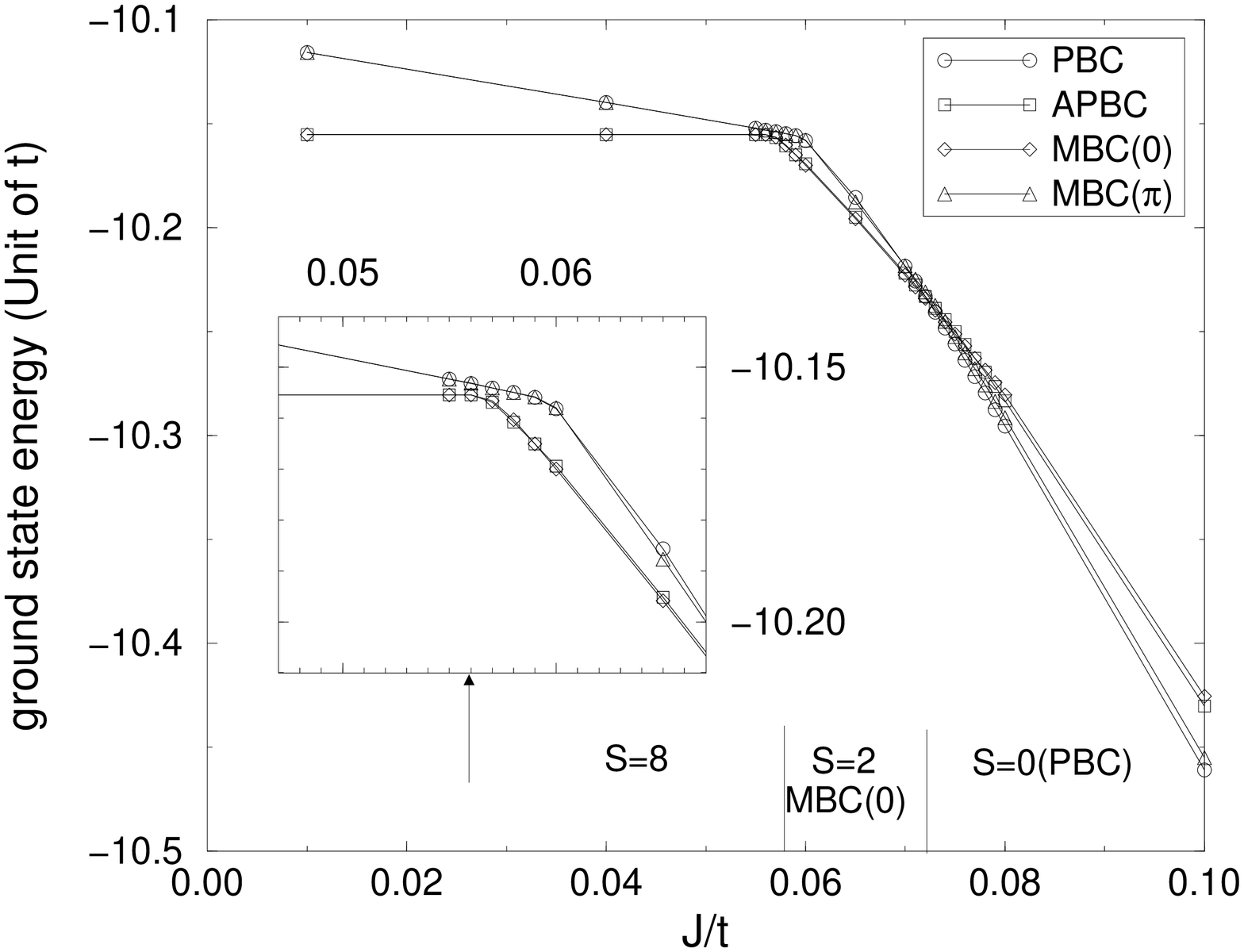}}}
\vspace{1mm}
\caption{
The upper(lower) graphs show the data for a 5(10)-rung ladder 
with 2(4) holes for different BC. The insets show
the magnified region where the saturated ferromagnet appears.
The
total spin of the 
ground state for the 10-rung ladder is indicated  
at the bottom of the lower graphs.}
\label{ferrened2} 
\end{figure}

	In Fig.~\ref{ferrened1} and~\ref{ferrened2} the ground state energy
is shown for different cases. In Fig.~\ref{ferrened1} the upper graph shows
the lowest energies for a 
5-rung ladder with one hole and the lower graph for a 10-rungs ladder
with two holes.
In Fig.~\ref{ferrened2} the lowest energies for a
5-rung ladder with two holes and for a 10-rung ladder with four
holes is shown in the upper and lower graph, respectively.
Each curve corresponds to different BC. For each
graph, the transition to the saturated ferromagnet where the energy is
$J/t$-independent is clearly seen for CSBC.
The two other BC do not have CS and the corresponding
energy states are singlets showing
a linear $J/t$-dependent energy lying above the
energy of the fully polarized state. This case will not be considered
further. 

	A crude finite size scaling can be made by extrapolating the critical
values of a 5-rung and a 10-rung ladder\cite{HaywardPoil}.
	At a doping of $\delta =0.1$, for $L=5$ the lowest critical 
value is given by APBC with value \mbox{$(J/t)_{5}\simeq0.052$}
while for $L=10$ both APBC and MBC($0$) gives the
same critical value of \mbox{$(J/t)_{10}\simeq 0.042$}. The infinite
extrapolation gives  
$(J/t)_{\infty}\simeq 0.032$. For \mbox{$\delta=0.2$}, 
the two values are
$(J/t)_{5}\simeq 0.078$ and \mbox{$(J/t)_{10}\simeq 0.056$} respectively, and the
infinite extrapolation gives \mbox{$(J/t)_{\infty}\simeq 0.034$}.
	For higher doping, i.e. for a 5-rung ladder with three holes and
for a 10-rung ladder with five and six  holes with  \mbox{$J/t \ge 0$}
no ferromagnetic ground state has been found. 

	In conclusion the diagonalization of small clusters shows signs 
of saturated ferromagnetism up to a doping of \mbox{$\delta \simeq 0.2$} and
for small 
values of \mbox{$J/t$} (\mbox{$J/t \simeq 0.033$}). 

\subsection{For $J/t>(J/t)_{\textrm{f}}$, Sign of Partial Ferromagnetism?}

For some clusters, the transition from the Nagaoka phase to the singlet
phase does not occur immediately, but passes through different phases
with an intermediate spin \mbox{$0<S<N_e/2$}. 
The lowest energy of the \mbox{$n_{\mathrm{h}}=2$}, $10$-rung ladder for
\mbox{$J/t<0.09$} is always given by APBC (with the exception of a very tiny
region close to \mbox{$J/t=0.04$}). In that  
region, the spin of the corresponding ground state passes through the finite values
$ S=5$ for \mbox{$0.043\leq J/t\leq 0.057$} and $S=1$ for \mbox{$0.058\leq
J/t\leq 0.062 $}. For \mbox{$J/t>0.62$} the ground state is a singlet with a
finite momentum of $k=\pi/5$ and a singlet at zero momentum for 
\mbox{$J/t\geq 0.083$}; they are mentioned at the bottom of Fig.~\ref{ferrened1}. 
For a \mbox{10-rung} ladder
with \mbox{$n_{\mathrm{h}}=4$} (Fig.~\ref{ferrened2}), some partial
ferromagnetism also occur. For \mbox{$0.057 \leq J/t \leq 0.072$} the spin is
\mbox{$S=2$} with MBC$(0)$ giving the lowest energy. 
For \mbox{$J/t> 0.072$}, the lowest energy is given by PBC at \mbox{$k_x=0$}
and is a singlet.

\begin{figure}
\centerline{\rotatebox{-90}{\epsfysize=9.cm\epsfbox{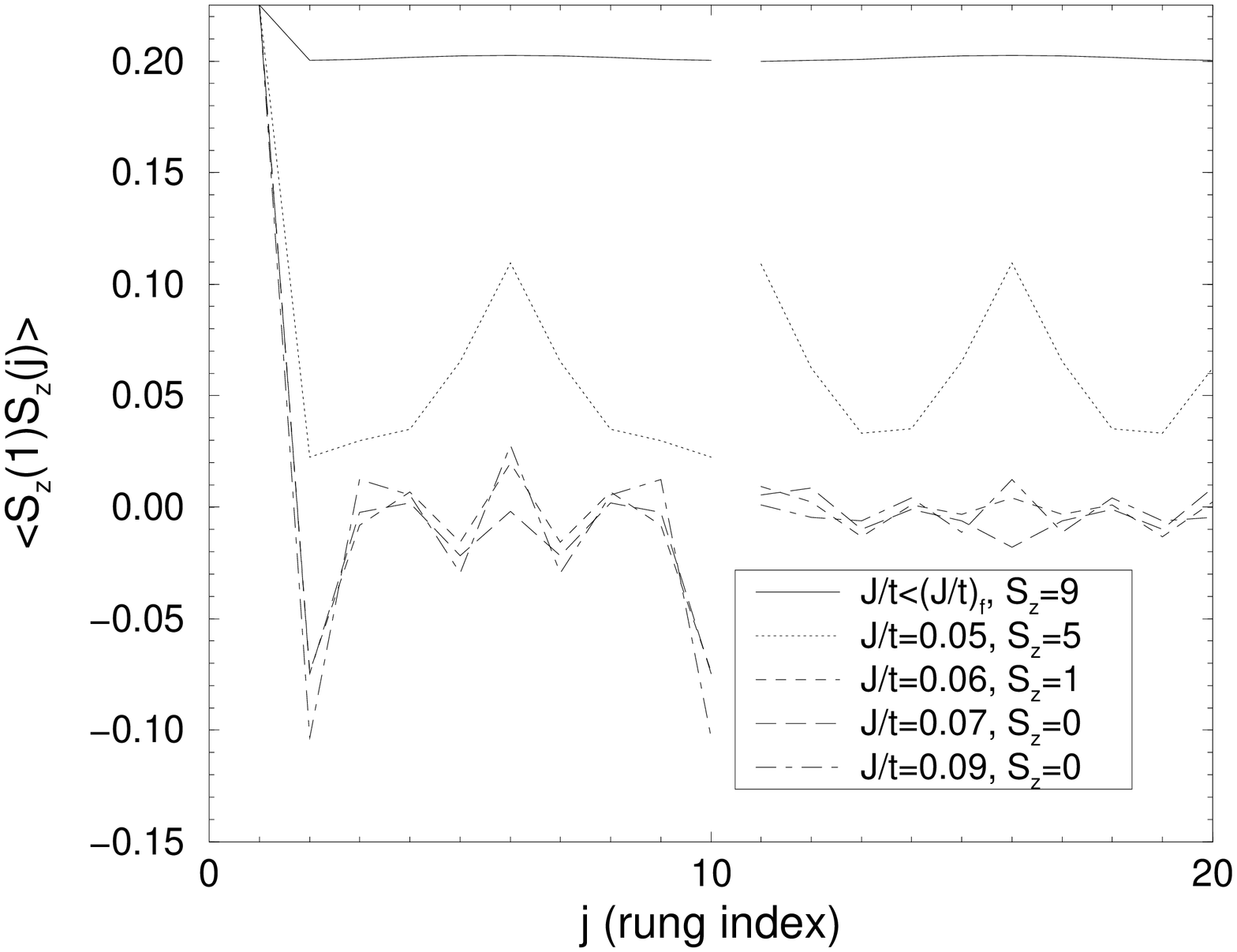}}}
\centerline{\rotatebox{-90}{\epsfysize=9.cm\epsfbox{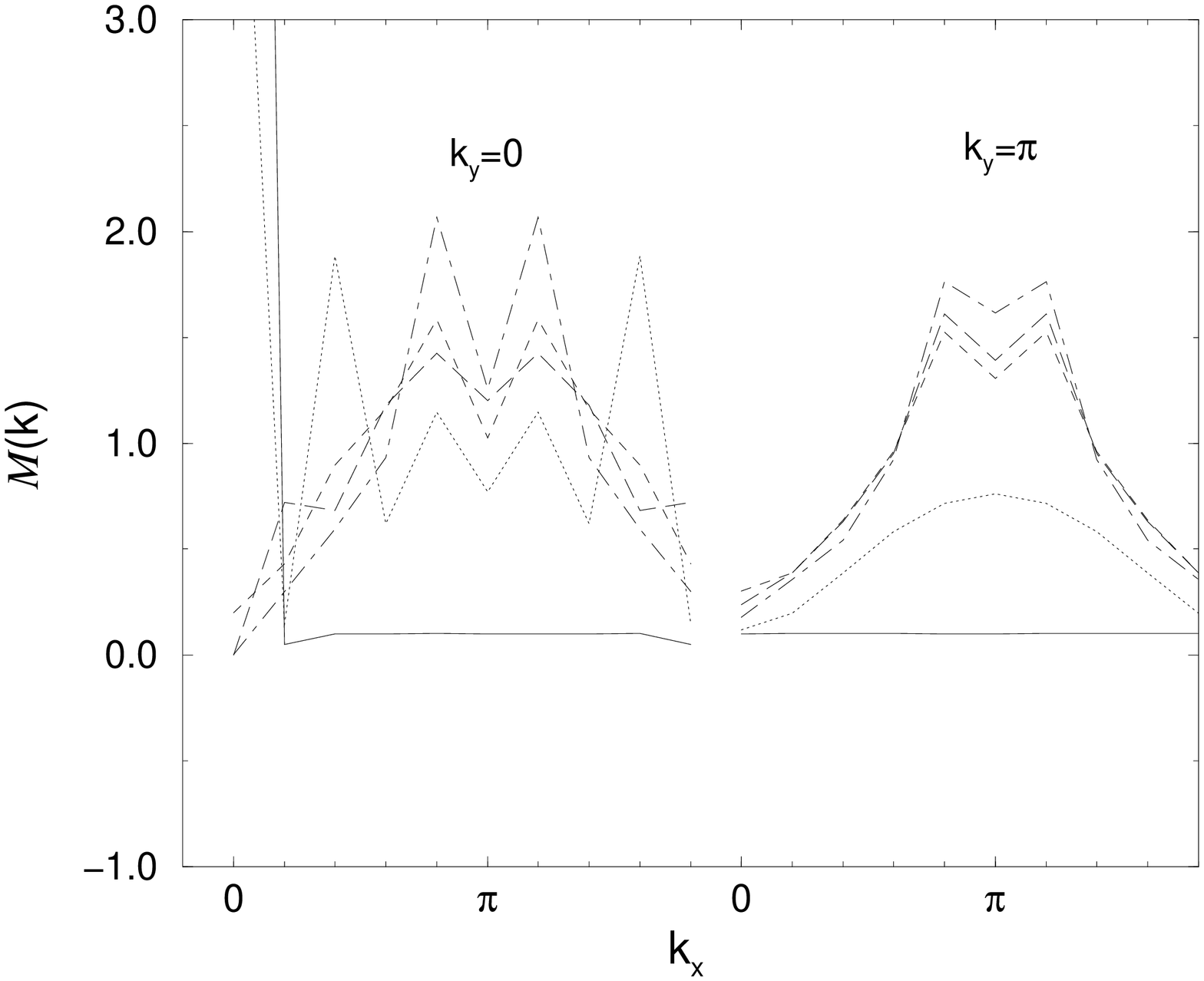}}}
\vspace{1mm}
\caption{The spin--spin correlation for a 10-rung ladder doped with two holes 
in real space and its respective Fourier transform 
at small values of $J/t$ for $S_z$=$S$ and using APBC. The 
$\cSk(\bk=0)\propto \expect{S_z^2}$ are out of the figure for  
the $S_z=9$, and $S_z=5$ cases.}
\label{ssl}
\end{figure}

The spin--spin correlations for the 10-rung ladder with $n_h=2$
and for $S_z=S$ are plotted in
Fig.~\ref{ssl}.
The correlations in real space are plotted in the uppermost graph
where the sites  on the first leg are
labeled \mbox{$1\leq j \leq 10$}, and on the second leg \mbox{$11\leq j \leq
20$} as pictured in Fig.~\ref{mbc10h2}. For $J/t\geq 0.05$, 
they show an alternating behavior around a ferromagnetic value 
indicating antiferromagnetic correlations. 
However, their Fourier transform, plotted in the lower graph, 
do not clearly indicate a continuous process. 
For \mbox{$J/t=0.07$} the correlation functions
have a  maximum in the branch \mbox{$k_y=\pi$} indicating
that the sum of the interband scattering processes is greater than the
intraband processes and thus, that the correlations between 
both bands are important.  The correlations at other values of $J/t$ 
display a maximum in the \mbox{$k_y=0$} branch showing that
the intraband processes are now favored.

 These effects are not found for small systems (L=5). It is thus 
tempting to conclude the existence of narrow region with
partial ferromagnetism phase in the thermodynamic
limit of the \mbox{$n$--$J$} phase diagram. However the present results are
not enough to draw a clear conclusion on this subject and the existence
of these phases is still open.

\section{Existence of a C2S2 Region?}
\label{c2s2}

The spin gap of 
 a doped system is intrinsically bound to the
formation of hole pairs. Moreover, it is known that
for very low \mbox{$J/t$}, when the system is ferromagnetic, holes repel
each other. The question of whether the transition between repulsive
and attractive holes occurs at the ferromagnetic to paramagnetic transition will
be addressed in this section. Numerical evidence shows that a region of
holes with repulsive residual interactions, 
and thus with gapless spin excitations, occurs between the
ferromagnetic and the LE phase. 

\subsection{Hole--Hole Correlations}

\begin{figure}[p]
\centerline{\rotatebox{-90}{\epsfysize=9.cm\epsfbox{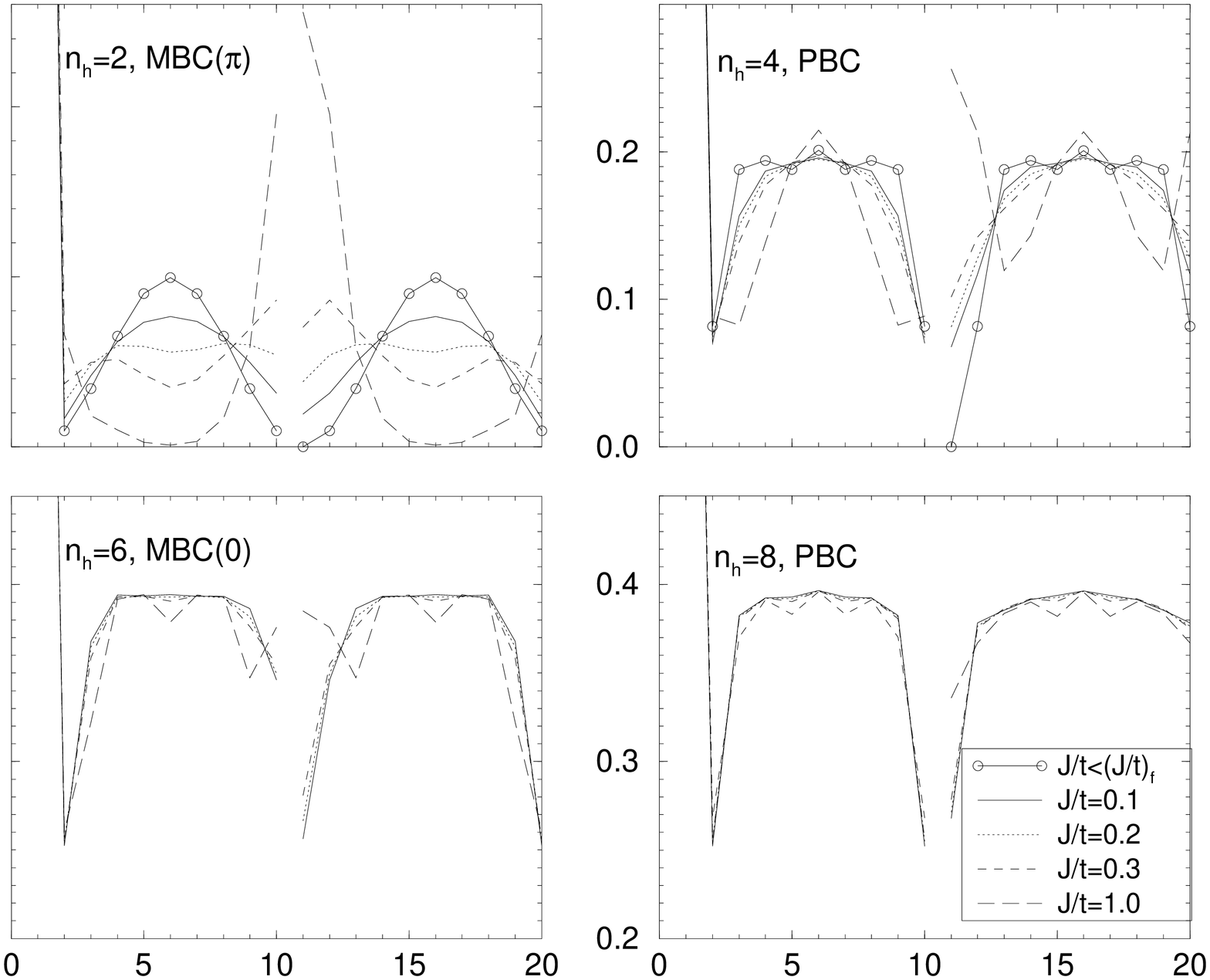}}}
\vspace{1.3cm}
\centerline{\rotatebox{-90}{\epsfysize=9.cm\epsfbox{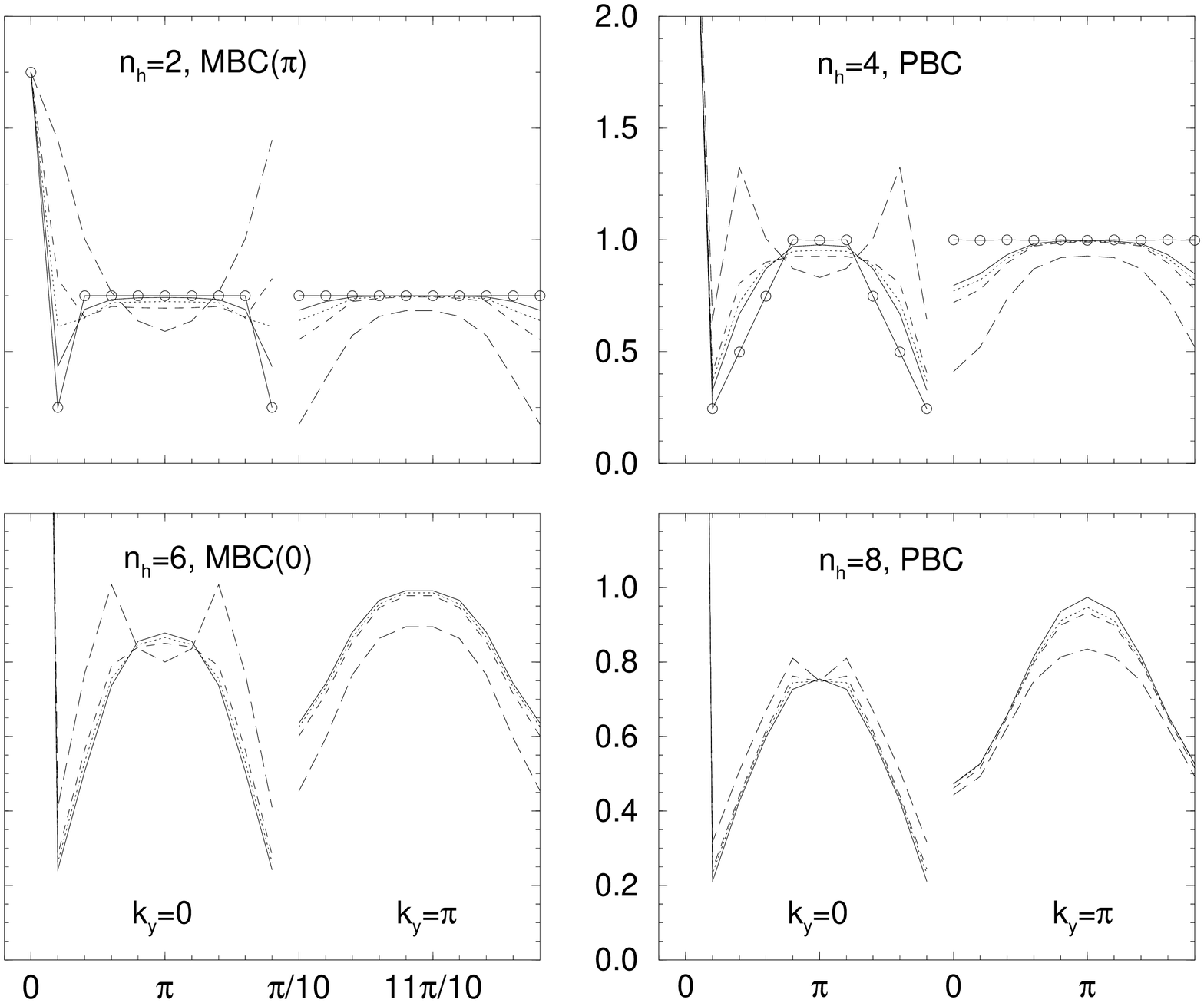}}}
\vspace{1.5cm}
\caption{The instantaneous hole--hole correlations for
a 10-rung isotropic ladder at different doping. The upper graphs give the
correlation in the site representation while the lower graphs
show their respective Fourier transforms. The convention of the labeling is
that of Fig.~\ref{mbc10h2} and~\ref{Fourier}.}
\label{corrh}
\end{figure}

A first insight into this question can be obtained by looking at the 
hole--hole correlation function \mbox{$\expect{n_{\mathrm{h}}(1)n_{\mathrm{h}}(j)}$}.
In Fig.~\ref{corrh} the correlations are
plotted for different values of \mbox{$J/t$} and 
different fillings, namely \mbox{$n_{\mathrm{h}}=2$} with MBC$(\pi)$,
$n_{\mathrm{h}}=4$ with PBC, \mbox{$n_h=6$} with MBC$(0)$ and \mbox{$n_{\mathrm{h}}=8$} with
PBC \cite{specialcare}.  
For \mbox{$n_{\mathrm{h}}=2$} and \mbox{$n_{\mathrm{h}}=4$} the correlation for the
ferromagnetic state 
computed with APBC is
also shown (circles). 
The uppermost graphs in the figure show the correlation functions in real
space with the site convention pictured in Fig.~\ref{mbc10h2}.
Each correlation is normalized such that
$\expect{n_{\mathrm{h}}(1)n_{\mathrm{h}}(1)}=1$.    

The lower graphs show the Fourier transform
\begin{equation}
\expect{n_{\mathrm{h}}(k)n_{\mathrm{h}}(-k)} 
\propto \cNk(\bk) \equiv \sum_{\bj}
e^{i\bk \bj} \expect{n_{\mathrm{h}}(1)n_{\mathrm{h}}(\bj)},
\end{equation}
 where $\bk$ is defined in Sec.~\ref{bouncon}.
The normalization corresponds to \mbox{$\cNk(0)=n_{\mathrm{h}}$} (out of the
graphs).  
In Fourier space, the points in the \mbox{$k_y=0(\pi)$} branch
are given by the left(right) curves in each graph. 

In real space, each graph shows the same behavior. 
 For instance 
for \mbox{$n_h=2(\delta=0.1)$} at \mbox{$J/t=0.1$} the second hole is found to
sit at the 
farthest point from the first hole. When \mbox{$J/t$} is
increased the hole--hole correlation for having two holes on the same
rung increases. When \mbox{$J/t=1$} the hole--hole correlation clearly shows
their bound character.

In the large \mbox{$\Jperp, \tperp$} limit, with \mbox{$\Jperp>2\tperp$},
the hopping of hole pairs between rungs is given 
in second order perturbation theory by 
\mbox{$\widetilde{t_{||}}=-2 t_{||}^2 /(\Jperp -2 \tperp)$}.
As the hole pairs act as hard-core bosons, they repel 
 each other in order to gain the maximal kinetic
energy. This can be viewed in the lower graphs plotting the
Fourier-transform, where peak in the \mbox{$k_y=0$} branch appear. This feature
clearly appears in the four different graphs. At the same time, the total
weight in the \mbox{$k_y=\pi$} decreases. 

This behavior 
raises the question of the existence of an
intermediate phase below the LE phase. 
In fact, for two holes and values of  \mbox{$0.05 <J/t<0.2$} the
system seems to be in a phase where holes repel each other. According
to the above picture, it would imply  spin-gapless excitations. 
However, it could also be a pure finite size effect in  
 that the radius of a two-hole bound state is greater than the length  of the 
ladder sample.
Thus, finite size effects strongly complicate the interpretation.
 The above correlations are not enough to conclude definitively to 
the existence of a new phase. 

\subsection{Spin--Spin Correlations}
\label{sec:sc}
\begin{figure}[p]
\centerline{\rotatebox{-90}{\epsfysize=9.cm\epsfbox{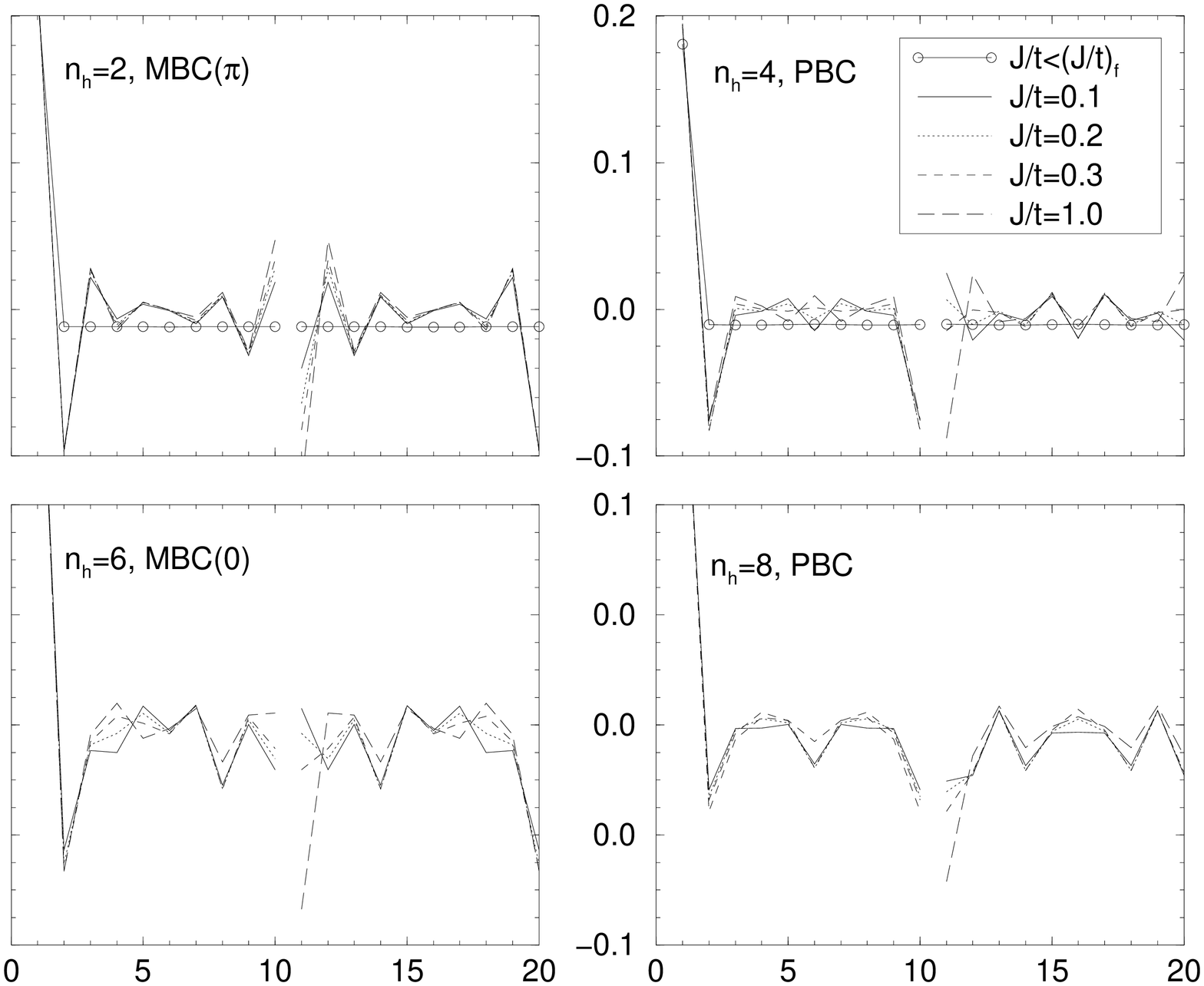}}}
\vspace{1.3cm}
\centerline{\rotatebox{-90}{\epsfysize=9.cm\epsfbox{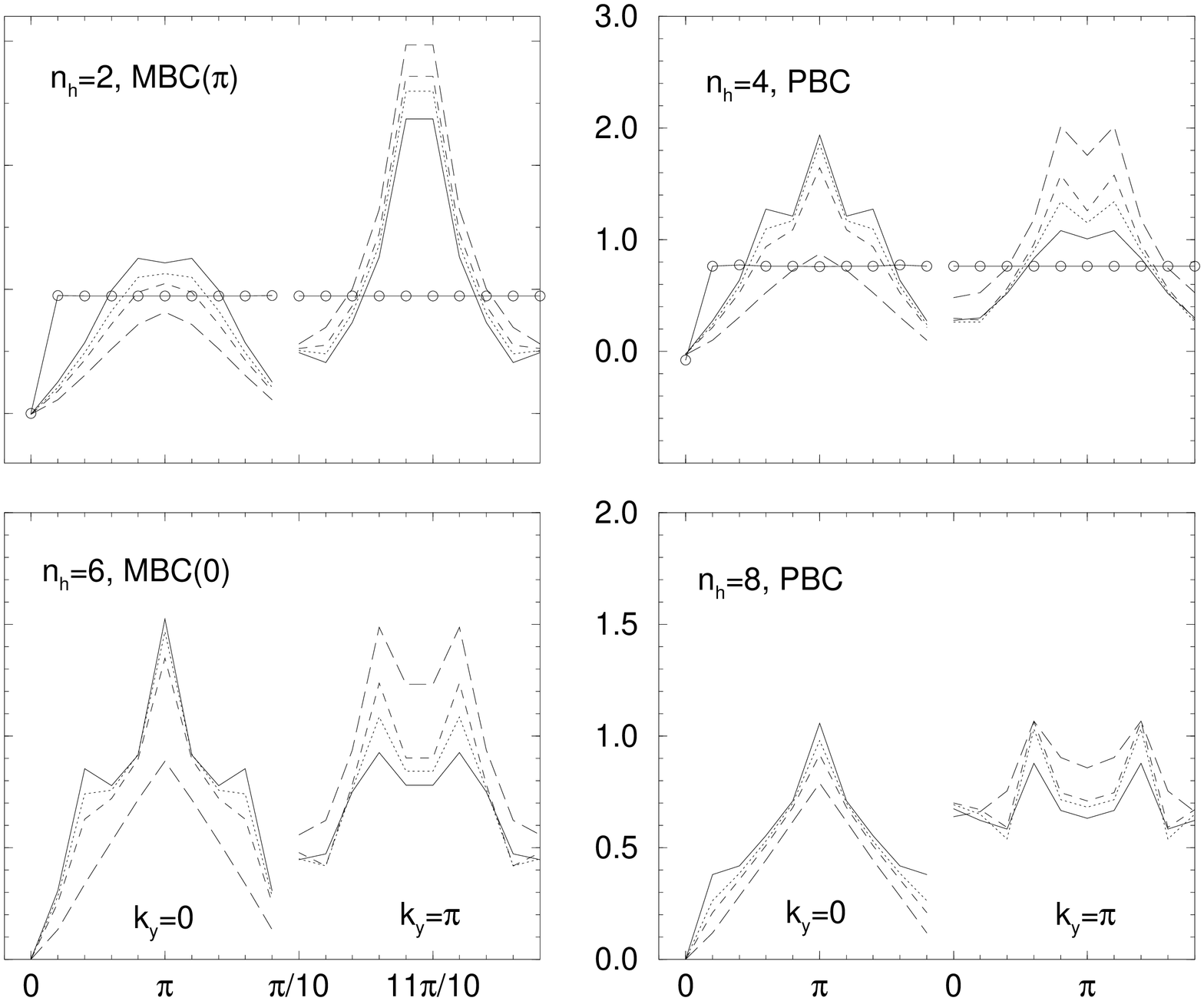}}}
\vspace{15mm}
\caption{The instantaneous spin--spin correlations  for
a 10-rung isotropic ladder at different doping. The upper graphs give the
correlations in the site representation while the lower graphs
show their respective Fourier transforms. The convention of the labeling is
the one of Fig.~\ref{mbc10h2} and~\ref{Fourier}.}
\label{corrs} 
\end{figure}

The spin--spin correlation of the systems
are plotted in Fig.~\ref{corrs}.
The uppermost graphs show the spin-spin correlations
$\expect{S_z(1)S_z(j)}$ in real space. They clearly show the 
antiferromagnetic  ordering of the spins along the chain. For
$n_{\mathrm{h}}=4$ and  
$n_{\mathrm{h}}=6$ and for low value of $J/t$, the correlation across the rung 
$\expect{S_z(1)S_z(11)}$
is ferromagnetic not antiferromagnetic. Increasing $J/t$ however
stabilizes the system to have antiferromagnetic ordering across the
rung. 

 This feature is also clearly emphasized by the Fourier transform,
plotted in the lower graph of Fig~\ref{corrs},
which have their maxima in the $k_y=0$ branch. Their Fourier transforms
$\cSk(\bk)$ also gives information on the different
scattering processes occurring between the different spins in the band
picture. In fact, they are proportional to the equal-time correlation
function,
\begin{equation}
	\cSk(\bk) \propto
\expect{S_z(k)S_z(-k)}=\sum_{\scriptsize\ket{\alpha}}|\mat{\alpha}{S_z(k)}{0}|^2,
\label{Szksum}
\end{equation}
where the sum is taken over all eigenstates of $H$ and where $\ket{0}$
denotes the ground state. 
For the case \mbox{$n_{\mathrm{h}}=2$}, the MBC($\pi$) has bonding
states filled up to \mbox{$k_{\mathrm{F,b}}=\frac{\pi}{2}$} and antibonding
 to \mbox{$k_{\mathrm{F,a}}=\frac{\pi}{5}$} as schematically shown in 
Fig.~\ref{Closedshell}. The lowest energy intraband excitations 
in the bonding and antibonding band occur at
\mbox{$\bk_b=(2k_{\mathrm{F,b}}+\frac{2\pi}{L},0)=(\frac{6\pi}{5},0)$} and
at \mbox{$\bk_a=(2k_{\mathrm{F,a}}+\frac{2\pi}{L},0)=(\frac{3\pi}{5},0)$},
respectively.
The corresponding curves in Fig.~\ref{corrs} show a maximum at $\bk_b$
while no peak appear at $\bk_a$ showing that these processes do not 
dominate in the sum of Eq.~\ref{Szksum}. 
Moreover, these processes are quite small compared 
to the interband processes plotted in the right part of the graph. According
to the band picture the lowest energy interband processes are at
\mbox{$\bk_{ab}=(k_{\mathrm{F,b}}+k_{\mathrm{F,a}}+\frac{\pi}{L}=
\frac{9\pi}{10},\pi)$.} 
 The curve display a maximum at that point dominating all the
other processes and especially the intraband scattering ones. 
It leads to the interpretation that the bonding and antibonding 
particles are strongly correlated at $E_F$. In the next 
section a picture will be proposed in which the gapless phase is a
manifestation of the absence of correlation between  bands. In such a picture,
the 
correlation for  \mbox{$n_{\mathrm{h}}=2$} is characteristic for a gapped
phase.

For \mbox{$n_{\mathrm{h}}=4$}(PBC) the band picture in the non interacting
limit 
predicts dominant intra-bonding band scattering processes at
\mbox{$\bk_b=(\pi,0)$}   
and intra-antibonding band scattering processes at \mbox{$\bk_a=(3\pi/5,0)$}.
Dominant interband scattering processes occur at
\mbox{$\bk_{ab}=(4\pi/5,\pi)$}.  
These values correspond to the (local)maxima of the correlation in
agreement with the simple non-interacting band picture.

For \mbox{$n_{\mathrm{h}}=6$}(MBC($0$)) the dominant scattering processes
are  expected at 
\mbox{$\bk_b=(\pi,0)$}, \mbox{$\bk_a=(2\pi/5,0)$}, and
 \mbox{$\bk_{ab}=(7\pi/10,\pi)$}.
For \mbox{$n_{\mathrm{h}}=8$}(PBC) the dominant scattering processes are  expected at 
\mbox{$\bk_b=(\pi,0)$}, \mbox{$\bk_a=(\pi/5,0)$}, and at
$\bk_{ab}=(3\pi/5,\pi)$. The corresponding graphs display corresponding
maxima in good agreement with the band-picture predictions.

\begin{figure}
\centerline{\rotatebox{-90}{\epsfysize=9.cm\epsfbox{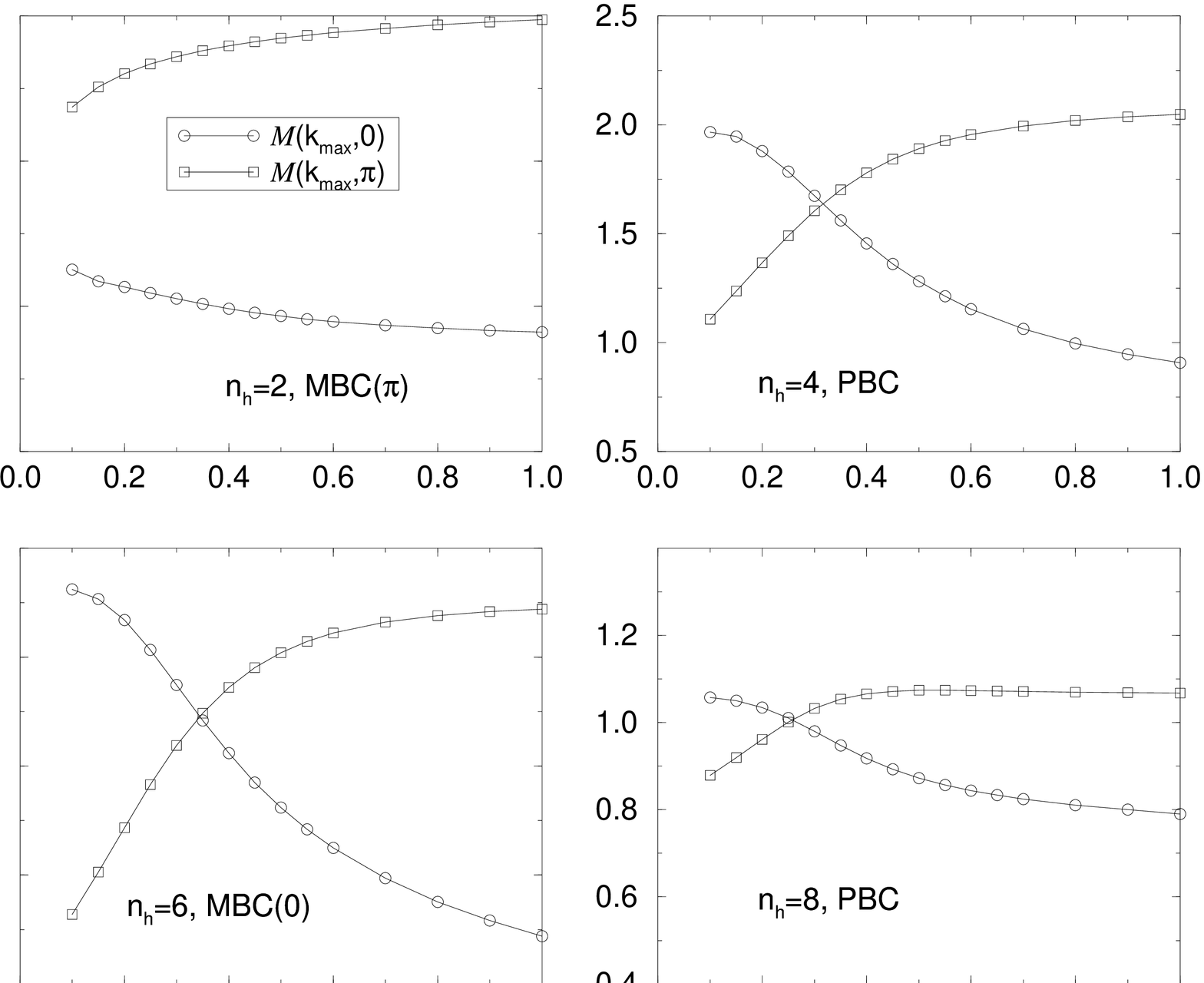}}}
\vspace{15mm}
\caption{10-rung ladder at different fillings.
The graphs show the maxima of the Fourier transform $\cSk(\bk)$
in the branches $k_y=0, \pi$. They 
correspond to excitations at $E_F$. The corresponding $k_x$
values are given in the text.} 
\label{corrsmax} 
\end{figure}

In Fig.~\ref{corrsmax} the maxima of the correlation functions for the 
$k_y=0$(circle) and the \mbox{$k_y=\pi$}(square) branches are plotted. 
For \mbox{$n_{\mathrm{h}}=4,6$} and \mbox{$n_{\mathrm{h}}=8$} 
they show a crossover from a region with dominant intraband scattering
to a region with dominant interband scattering. This suggests that
the system has uncorrelated bonding and antibonding bands at $E_F$ such that
the low energy physics 
is analogous to that of two one-dimensional
systems, with spin and charge gapless excitations (C2S2). 
A rough criteria for the transition can be defined by taking 
the critical $J/t$ at the crossover. This yields
\mbox{$(J/t)_{\mathbf c} \simeq 0.32,0.34,0.26$}
for \mbox{$\delta=0.2,0.3$}, and $\delta=0.4$, respectively.
For \mbox{$n_{\mathrm{h}}=2$} the interband
scattering is always much bigger than the intraband scattering yielding to
the conclusion that no C2S2 phase occurs for that particular filling.
This emphasizes that care must be taken in interpreting the simple hole--hole
correlations.

\subsection{(Anti-) Bonding Pair Correlations}
\label{sec:ssttc}

Defining the singlet \mbox{$\hat{S}^{\dag}_{i}$} and triplet
\mbox{$\hat{T}^{\dag}_{i0}$}, \mbox{$\hat{T}^{\dag}_{i\alpha}$}, $\alpha \in
\{\uparrow,\downarrow\}$  creation operator on the rung $i$ with
\begin{eqnarray}
	\hat{S}^{\dag}_{i} & = &  \frac{1}{\sqrt{2}}(\cd_{1i\uparrow}
	\cd_{2i\downarrow}- 
		\cd_{1i\downarrow}\cd_{2i\uparrow}),  \\
	\hat{T}^{\dag}_{i0} & = & \frac{1}{\sqrt{2}}(\cd_{1i\uparrow}
	\cd_{2i\downarrow}+
		\cd_{1i\downarrow}\cd_{2i\uparrow}),\\
	\hat{T}^{\dag}_{i\alpha} & = & \cd_{1i\alpha}\cd_{2i\alpha},
\end{eqnarray}
the different pair correlations between bonding and antibonding 
states
in the subspace of single occupied states can be written as
\cite{phasenote}
\begin{equation}
	\begin{array}{lcr}
	\expect{\bd_{i\uparrow} \bd_{i\downarrow} a_{j\uparrow} a_{j\downarrow}}
	& = & \frac{1}{2}\expect{\hat{S}^{\dag}_{i}\hat{S}_{j}}, \\[2mm]
	\expect{\bd_{i\uparrow} \ad_{i\downarrow} b_{j\uparrow}
	a_{j\downarrow}} & = & 
	-\frac{1}{2}\expect{\hat{T}^{\dag}_{i0}
	\hat{T}	_{j0}},\\[2mm]
	\expect{\bd_{i\alpha} \ad_{i\alpha} b_{j\beta} a_{j\beta}} & = &
	-\expect{\hat{T}^{\dag}_{i\alpha}
	\hat{T}	_{j\beta}}.
		\end{array}
\label{cbbaa2}
\end{equation}
In the C2S2 phase, when the bonding and antibonding states 
are not correlated, all pair correlations are shortrange.
In the spin-gapped region, the ground state is characterized by 
the formation of interchain singlet. Thus, long range singlet-singlet (SS)
pair correlations will appear while the triplet--triplet (TT) pair
correlations will remain short-range.

\begin{figure}
\centerline{\rotatebox{-90}{\epsfysize=9.cm\epsfbox{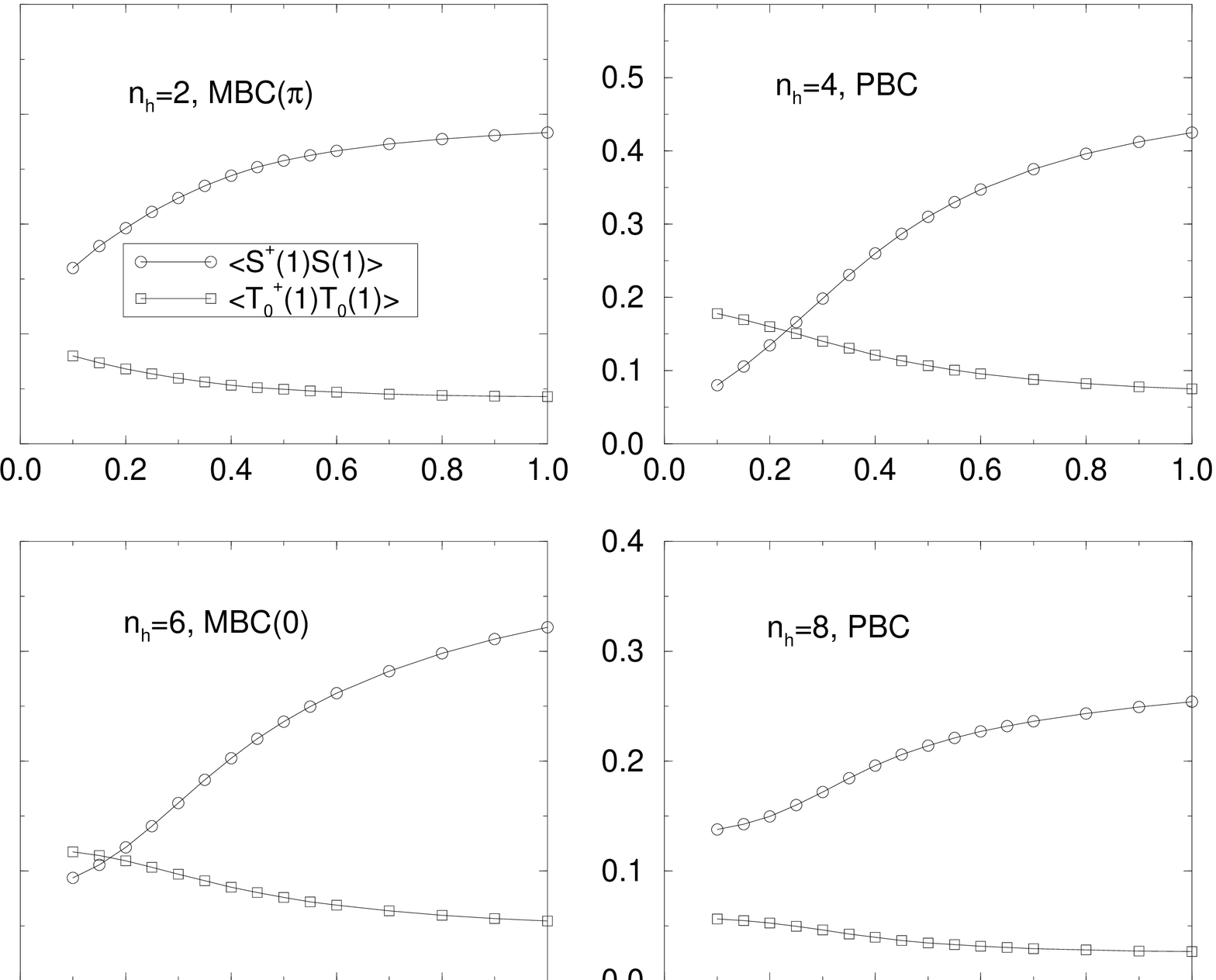}}}
\vspace{1.5cm}
\caption{The pair correlations $\SSe{1}{1}$, $\TTo{1}{1}$ as function of $J/t$ 
for  a 10-rung isotropic ladder at different doping.}
\label{st10h24s1s1}
\end{figure}

For an undoped system,  the pair correlations are finite only on the same
rung   $j = i$, where they are proportional to the number operator 
for singlet or triplet states,
and depend only on the ratio \mbox{$J_{\perp}/\Jpar$}. For \mbox{$\Jpar=0$},
the singlet--singlet  pair correlation at $j=i$ is  1  while the
triplet--triplet pair correlation at $j=i$ is 0. 
At the isotropic point, the values
for the undoped 10-rung ladder are 
\begin{eqnarray}
	\SSe{1}{1} & \simeq & 0.7, \\
	\TTo{1}{1} & = & \TTu{1}{1}  \simeq  0.1. 
\end{eqnarray}
In a doped system, the pair correlations at the isotropic point depend on  
the ratio $J/t$. 
	In Fig.~\ref{st10h24s1s1} \mbox{$\SSe{1}{1}$}
and \mbox{$\TTo{1}{1}$} are plotted as function of $J/t$. The case
\mbox{$n_{\mathrm{h}}=2$} 
shows that 
\mbox{$\SSe{1}{1}> \TTo{1}{1}$} for all plotted  values $J/t$. It favors
the scenario that a 10-rung ladder with two holes has a gap for
very low $J/t$-values in agreement with the discussion of the
spin--spin correlation. With increasing $J/t$ the singlet number
$\SSe{1}{1}$ rapidly increases
 while the triplet number \mbox{$\TTo{1}{1}$}  decreases. 
On contrary, for \mbox{$n_{\mathrm{h}}=4(6)$}
a cross-over between the singlet and triplet number occurs at
\mbox{$J/t \simeq 0.23(0.17)$}.
This is in agreement with the speculated gapless
region at low $J/t$.
The case with \mbox{$n_{\mathrm{h}}=8$},
exhibits a  favored rung singlet configuration. 

\begin{figure}
\centerline{\rotatebox{-90}{\epsfysize=9.cm\epsfbox{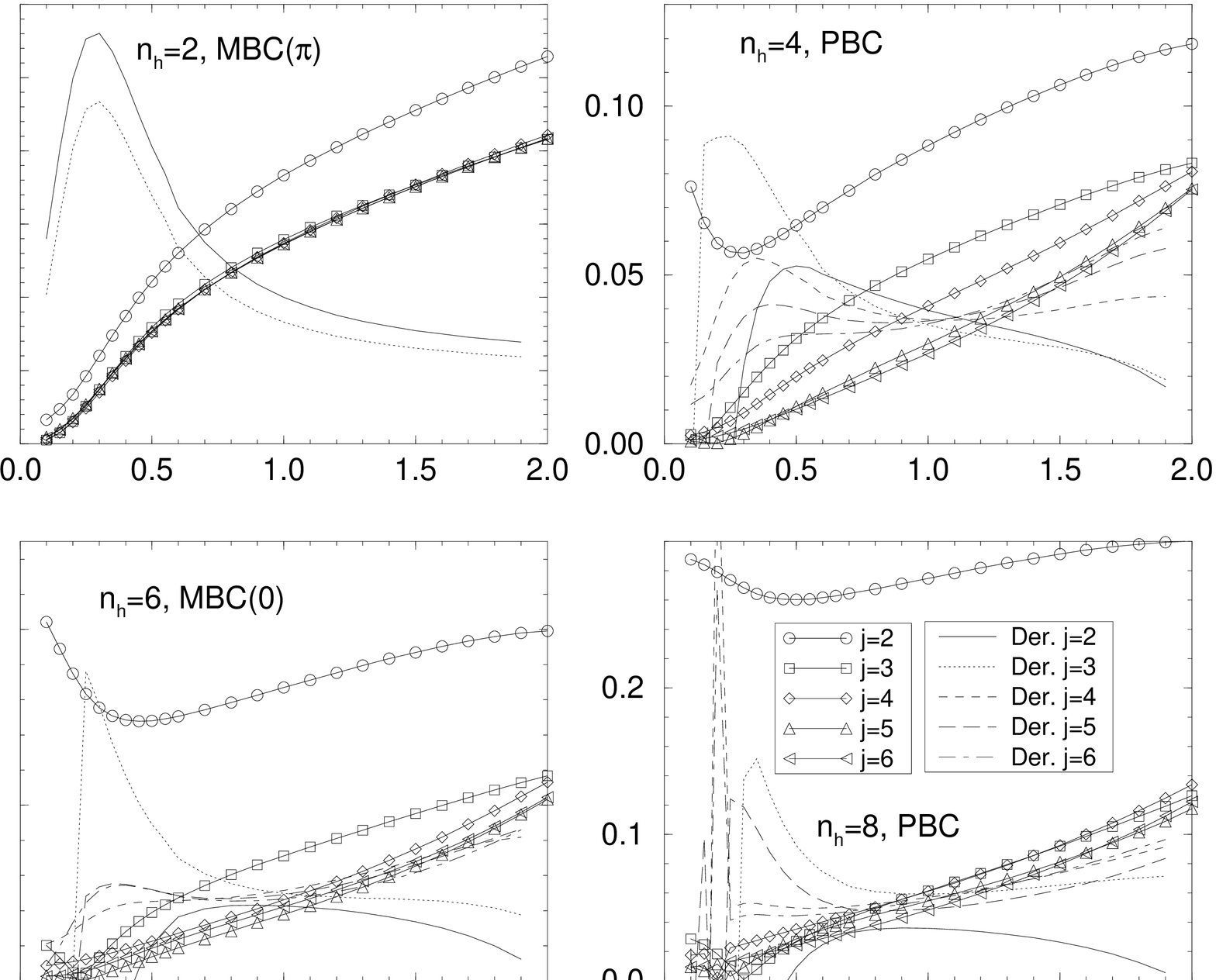}}}
\vspace{1.5cm}
\caption{The pair correlations $\SSe{1}{j}_r$ as function of $J/t$ 
for  a 10-rung isotropic ladder at different doping and for different
lattice site $j$.}
\label{st10h24s1s_6}
\end{figure}

	In Fig.~\ref{st10h24s1s_6} the pair correlation at different rungs is
shown 
as a function of $J/t$ for the different cases. To make a consistent 
comparison of the long-range order for various $J/t$ the pair correlations 
are normalized according to the singlet number, \mbox{$\SSe{1}{j}_r =
\SSe{1}{j}/\SSe{1}{1}$}. For \mbox{$n_{\mathrm{h}}=2$} the pair correlations uniformly
tend 
to zero when 
$J/t$ decreases to zero. The solid and dotted lines show the derivative
(forward difference) of the curves for $j=2$ and $j=3$. Both show a
peak around $J/t=0.3$ indicating the value of $J/t$ below which the
pair correlation become  short-ranged. The curve does not show any 
different behavior for $J/t>0.1$. Below this value, it has been seen that 
the system has a transition to a ferromagnetic phase at $J/t=0.032$. 
Between these two points some sign of partial ferromagnetism has been
discussed but no direct evidences for a C2S2 phases has been
observed.

The striking feature of the graphs for \mbox{$n_{\mathrm{h}}=4,6$} and
\mbox{$n_{\mathrm{h}}=8$} is the qualitatively different  behavior of
the different curves at small $J/t$. 
The normalized  
short range pair correlation \mbox{$\SSe{1}{2}_r$} shows a minimum at
\mbox{$J/t=0.30,0.45$}, and $0.50$ for \mbox{$n_{\mathrm{h}}=4,6$}, and $8$,
respectively.  The solid and dashed/dotted lines 
show the derivative of the different curves. Maxima in the
derivatives appear in both cases for \mbox{$0.25<J/t<0.5$} below which value
the  correlation get short range.

	A critical value will be defined to be at the point where the
long-range pair correlation \mbox{$\SSe{1}{5}_r$}	change the
slope of its 
curve. This gives the values
\mbox{$(\widetilde{J/t})_{\mathbf c} \simeq 0,20,0.25,0.25$}
for \mbox{$\delta=0.2,0.3$}, and $0.4$, respectively. 

\section{Transformed Hamiltonian}

 In this section, the physical picture for the C2S2 phase based on the absence
of phase correlation between the bonding and anti-bonding bands is examined.
For $J/t \simeq 0.5$ 
phase coherence between the bands at  $E_F$
exists leading to a LE liquid with the spingap. 
The question whether 
this phase coherence disappears for small $J/t$ leading to two independent 
LL (C2S2) will be investigated~\cite{Haldane}.
The terms in the Hamiltonian coupling both bands are of
the form
\mbox{$b_{k_1,\sigma}^{\dag}b_{k_2,\sigma'}^{\dag}a_{k_3,\sigma'}
a_{k_4,\sigma} + \hc$}. They must be irrelevant near $E_F$
in the gapless LL phase and become relevant when $J/t$ increases. 
To investigate this we transform the Hamiltonian by  introducing a relative phase between the bonding
and antibonding operators,
so that only terms of the above form are affected, then the dependence of 
the ground state on  this phase difference will be studied. 

\subsection{Relative Phase Transformation}

 The transformation is defined by
\begin{equation}
	H=PH(b,a)P \rightarrow PH(e^{i\phi_b}b,e^{i\phi_a}a)P =H_{\mathrm{t}},
\label{trans} 
\end{equation}
where  $P$ is the projector onto singly occupied states and
$H(a,b)$ is the $t$--$J$ Hamiltonian (without projection)
written in the bonding and  antibonding basis. 
The hopping term of the Hamiltonian is diagonal in the bonding and
antibonding basis and thus does not change.
The introduction of a phase shift will only affect 
terms of the form 
\[ \bd \bd a a \rightarrow  e^{i2(\phi_b -\phi_a)} \bd \bd a a, \] which 
occur in the
 \mbox{$H_{\Jpar}$} and \mbox{$H_{\Jperp}$} part.  They couple the bonding and
the 
antibonding operators and are thus responsible for the appearance of
the spin gap. 

With \mbox{$\gamma=\phi_b -\phi_a$}, The 
total transformation is summarized through the definition, 
\begin{equation}
 PH(c_{lj}) P \longrightarrow 
PH(c_{lj,\mathrm{t}}) P=H_{\mathrm{t}}(\gamma),
\label{transt}
\end{equation}
where the transformed Hamiltonian \mbox{$H_{\mathrm{t}}(\gamma)$} 
has a periodicity of $\pi$ in $\gamma$. 
The site operators are transformed according to the rule 
$c_{lj}  \longrightarrow  c_{lj,\mathrm{t}}$ with
\begin{eqnarray}
  c_{1j,\mathrm{t}} & = & \frac{1}{2}\left( (e^{i\phi_b}
+e^{i\phi_a})c_{1j} +(e^{i\phi_b}-e^{i\phi_a})c_{2j}  \right)
\label{transop1}\\ 
 c_{2j,\mathrm{t}} & = & \frac{1}{2}\left( (e^{i\phi_b}
-e^{i\phi_a})c_{1j}+(e^{i\phi_e}+e^{i\phi_a})c_{2j} \right).
\label{transop2}
\end{eqnarray}

The final transformed Hamiltonian can be written with three distinct terms
\[H_{\mathrm{t}}=H_{\mathrm{t},h}+H_{\mathrm{t,\Jpar}}+H_{\mathrm{t,\Jperp}},\]
where the first part $H_{\mathrm{t},h}$ is the hopping part being the same as
in the original Hamiltonian. 
The second term $H_{\mathrm{t,\Jpar}}$ involving magnetic coupling along the chain
contains four distinct terms
\begin{eqnarray}
H_{\mathrm{t},\Jpar}& = &
H_{\mathrm{t,\Jpar,1}}+H_{\mathrm{t,\Jpar,2}}
	  + H_{\mathrm{t,\Jpar,3}}+H_{\mathrm{t,\Jpar,4}}
\label{termpar}
\end{eqnarray}
which are listed and discussed in Appendix~\ref{appen}.  
The interleg magnetic coupling \mbox{$H_{\mathrm{t},\Jperp}$} vanishes
at 
\mbox{$\gamma=\pi$} (see Appendix~\ref{appen}) in agreement with the suppression of
correlation between the bonding and antibonding operators.

\subsection{Ground State Energy}

\begin{figure}
\centerline{\rotatebox{-90}{\epsfysize=9.cm\epsfbox{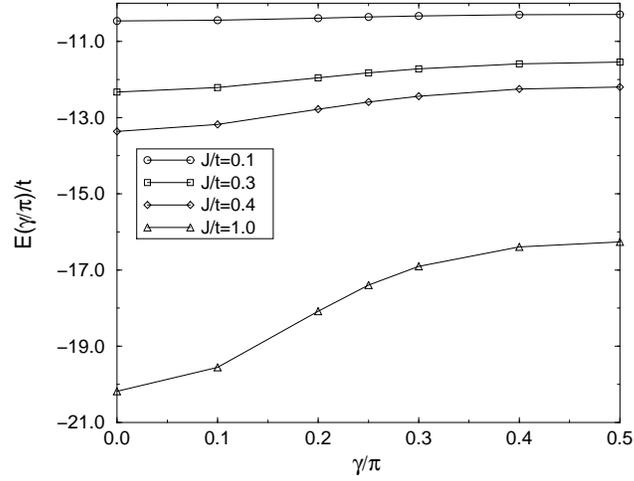}}}
\vspace{1mm}
\caption{The ground state energy vs.~$\gamma$ 
for a 10-rung ladder with 4 holes. The ground state energy is an
increasing function of $\gamma$.}
\label{gsgamma} 
\end{figure}

The ground state may not be completely decoupled into a
bonding and an antibonding part. Only the particles with a $k$-value near $E_F$
 must be uncorrelated for two independent LL. 
By calculating the ground state energy of $H_{\mathrm{t}}$
the influence of the phase shift on
all bonding and antibonding operators is included 
and it is thus possible that the influence of the phaseshift is 
qualitatively similar for different values of $J/t$. 
In fact, in Fig.~\ref{gsgamma} the ground state energy is plotted for a
10-rung ladder with 4 holes and using PBC. By switching on $\gamma$, the energy
 increases for all plotted values of $J/t$. However, no clear qualitative 
change in the 
evolution of the energy vs $\gamma$ for different $J/t$ can be observed. This
is a consequence of the fact that all particles (also away from  $E_F$) are
involved. 
	It is interesting to consider the different correlation
functions to look for signs if the  system undergoes a phase-transition when
$\gamma$ is altered. 

\subsection{Correlation Functions}

\begin{figure}
\centerline{\rotatebox{-90}{\epsfysize=9.cm\epsfbox{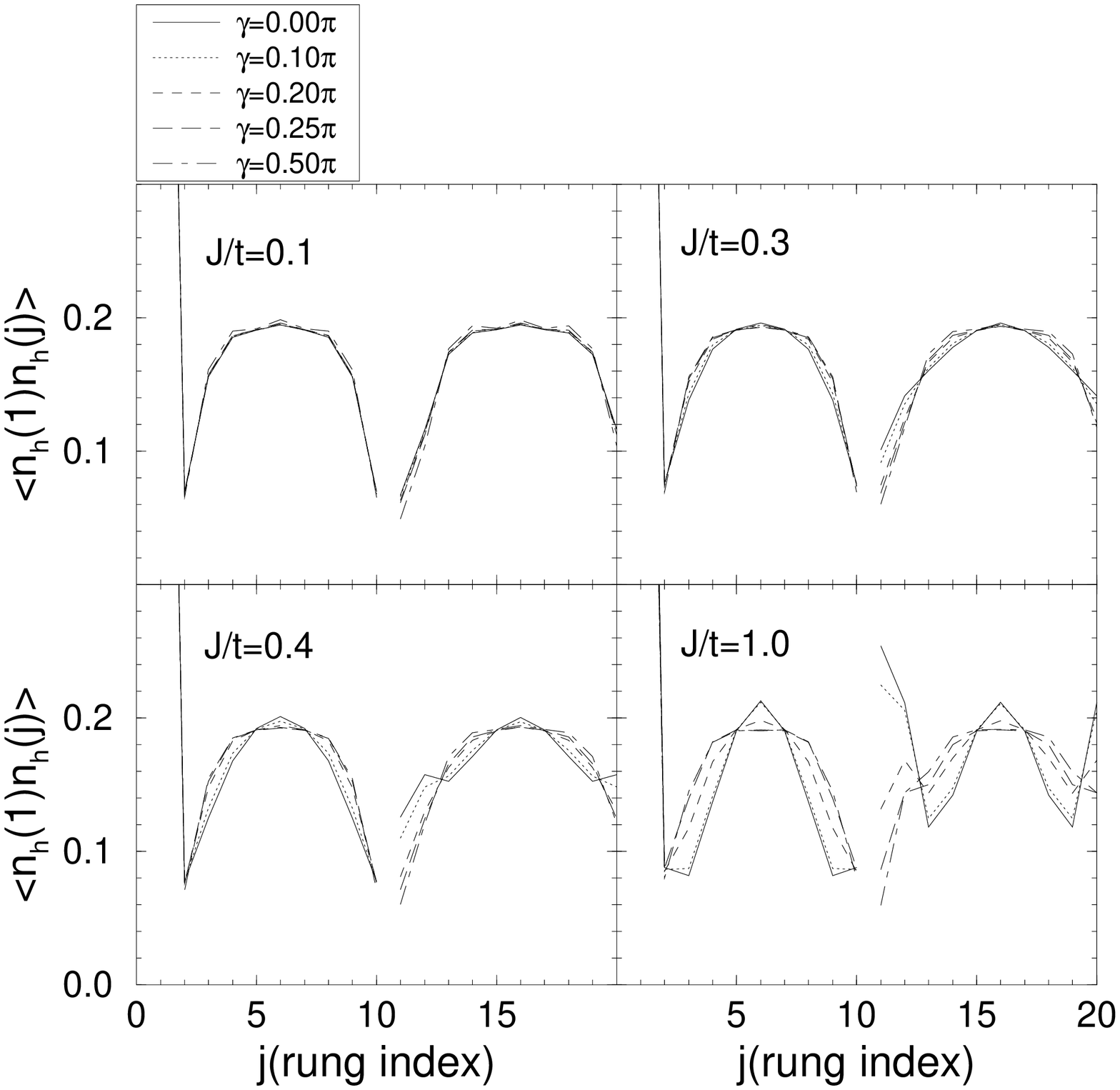}}}
\centerline{\rotatebox{-90}{\epsfysize=9.cm\epsfbox{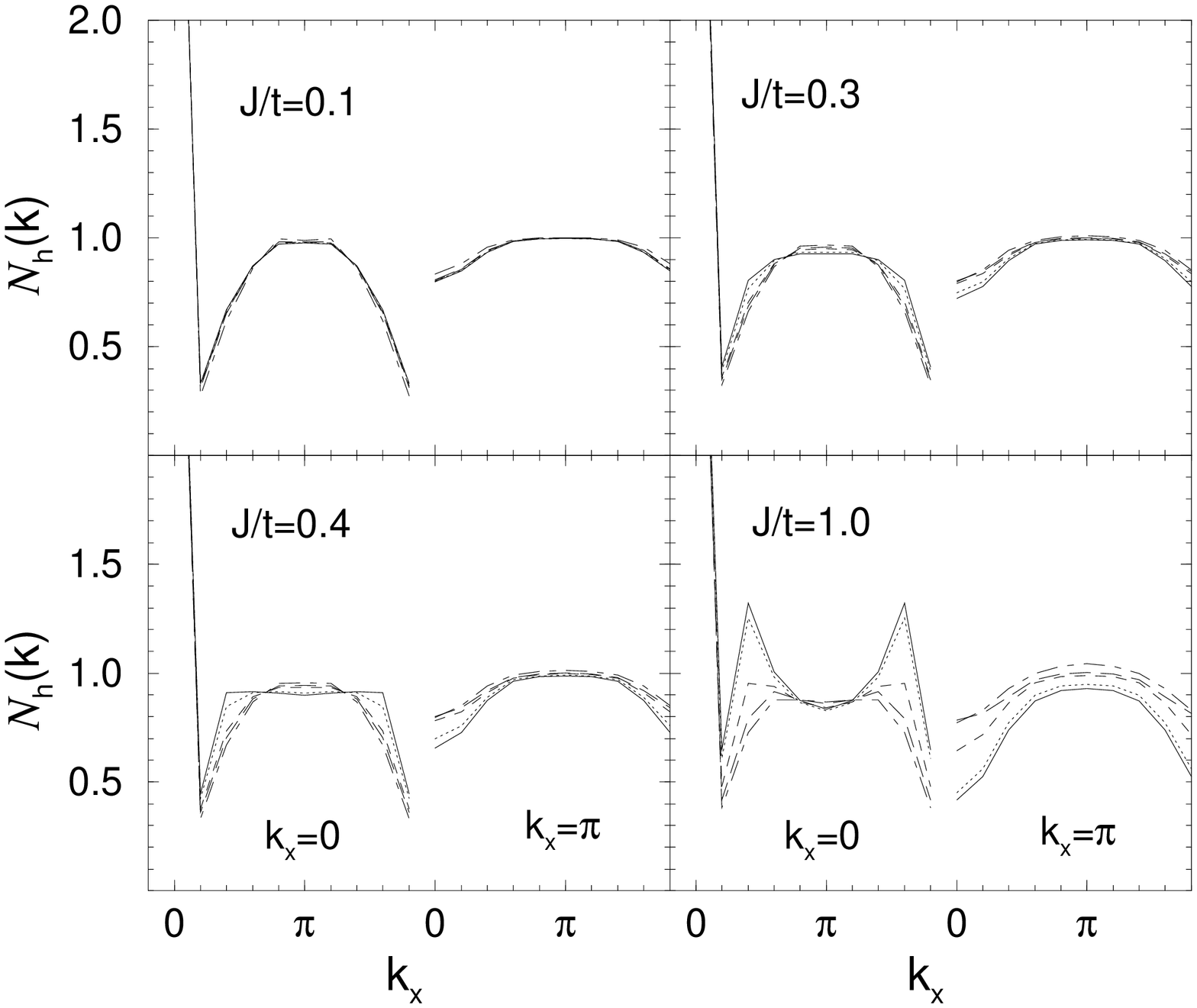}}}
\vspace{5mm}
\caption{The instantaneous hole--hole correlation of a
 10-rung isotropic ladder with 4 holes for different values of $J/t$ and
$\gamma$. The upper graphs show the correlation in the site representation
while the lower graphs show their Fourier transform. The convention of the
labeling is 
that of Fig.~\ref{mbc10h2} and~\ref{Fourier}.}
\label{corrhg}
\end{figure}

In Fig.~\ref{corrhg} the hole--hole correlations for a
10-rung ladder with 4 holes are plotted. The correlations are shown
at different $J/t$ for
different values of $\gamma$. The upper graphs show the correlations in
 real space, and the lower their Fourier transform.
For \mbox{$J/t \le 0.3$} only small changes in the shape of the correlations are
observed when $\gamma$ increases. For \mbox{$J/t=0.4$} and \mbox{$J/t=1.0$} the shape of the correlation is
clearly changed indicating state with repulsive interactions at
$\gamma=0.5\pi$. 
The influence of $\gamma$
for \mbox{$J/t=1.0$} is the most important. The rung correlation
$\expect{n_{\mathbf{h}}(1)n_{\mathbf{h}}(11)}$ is clearly suppressed in
agreement with the 
breaking of hole-pairs. The diagonal correlation
$\expect{n_{\mathbf{h}}(1)n_{\mathbf{h}}(12)}$ shows a peak at
\mbox{$\gamma=0.2\pi$} 
which is 
suppressed for \mbox{$\gamma=0.5\pi$. The peak in the branch $k_y=0$} of the 
Fourier transform,
characteristic of the homogeneity of the hole-pair repartition is
 also suppressed.

\begin{figure}
\centerline{\rotatebox{-90}{\epsfysize=9.cm\epsfbox{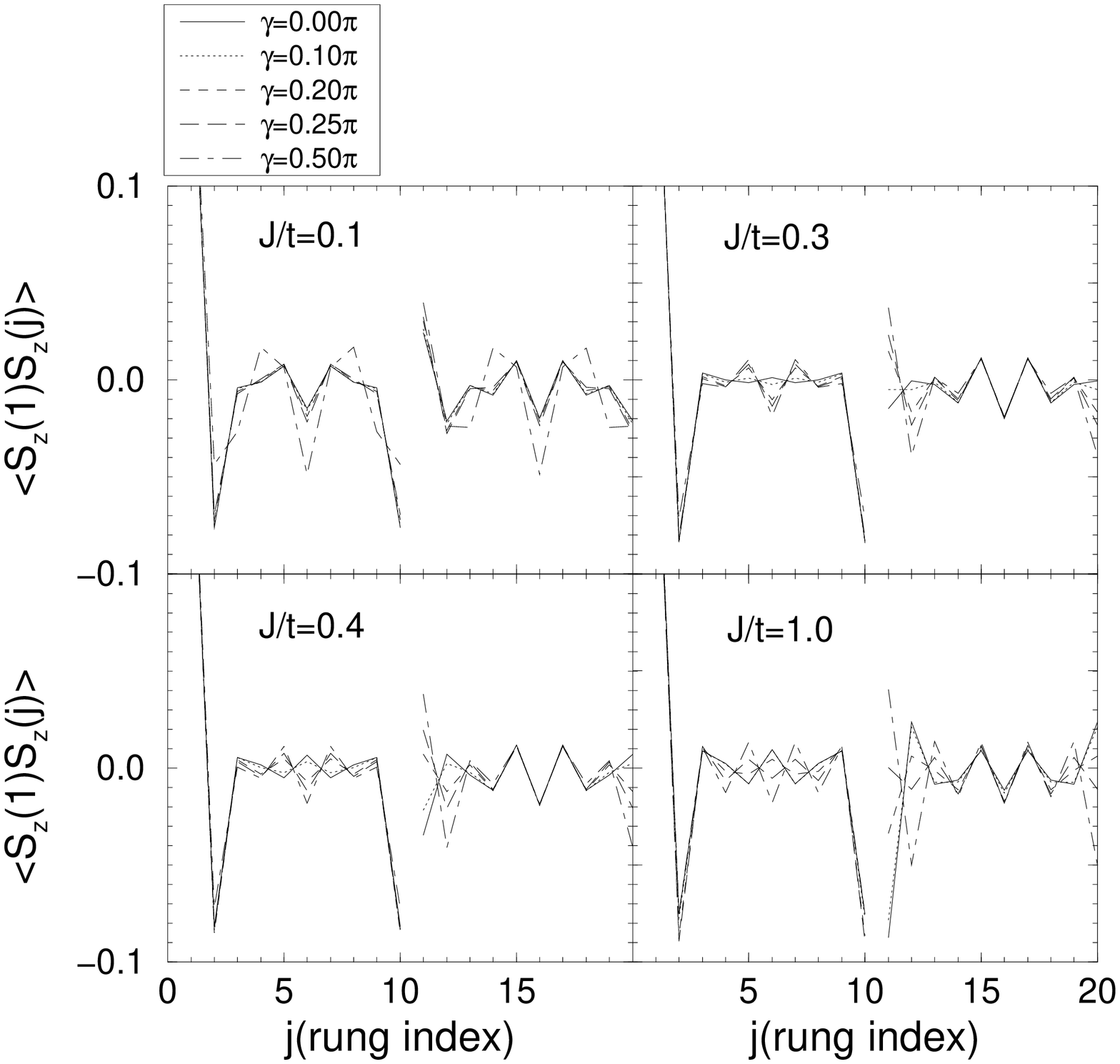}}}
\centerline{\rotatebox{-90}{\epsfysize=9.cm\epsfbox{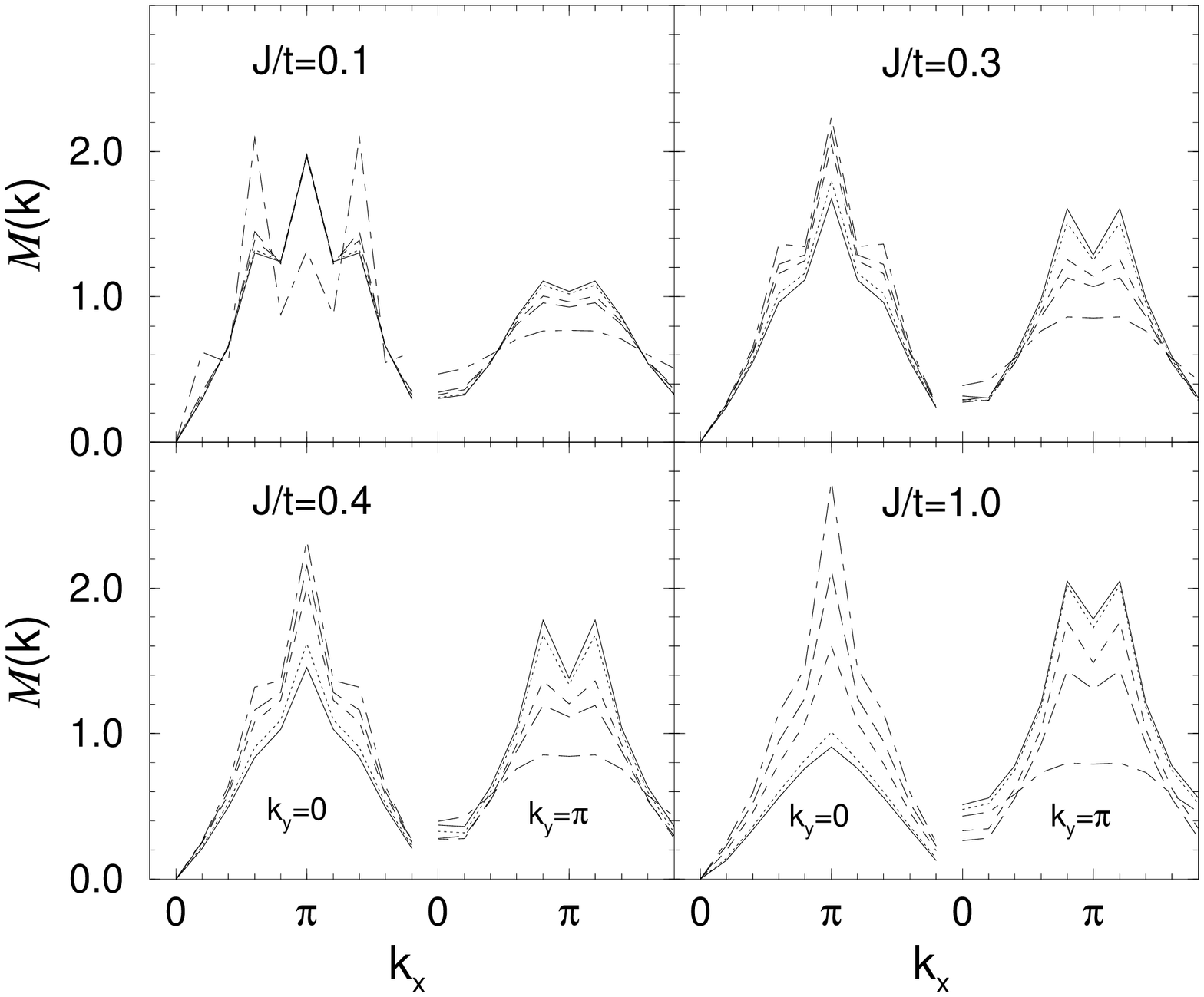}}}
\vspace{1mm}
\caption{The instantaneous spin--spin correlations  of
a 10-rung isotropic ladder with 4 holes for different values of $J/t$ and
$\gamma$. The upper graphs show the correlations in the site representation
while the lower graphs represent the Fourier transforms. The convention of the
labeling is 
that of Fig.~\ref{mbc10h2} and~\ref{Fourier}.}
\label{corrshg}
\end{figure}

	In Fig.~\ref{corrshg} the spin--spin correlation  for the
same cluster is plotted. The measured antiferromagnetism is quite sensitive to 
the phase shift. In  real space the antiferromagnetic correlation
across the rungs is turned into leg independent behavior 
(rung ferromagnetism). However, $\cSk(\bk)$ measures all spins 
in the band pictures, therefore not only those  at 
$E_F$, and a change in the correlation is also compatible 
with the gapless phase. 
The most striking picture of the spin--spin correlations is given by
their Fourier transforms. As discussed above, dominant peaks
in the bonding channel \mbox{($k_y=0$)} are expected for the gapless phase,
while dominant interband \mbox{($k_y=\pi$)} scattering processes are expected
in the gapped phase. This behavior appears clearly in all graphs. 

\begin{figure}
\centerline{\rotatebox{-90}{\epsfysize=9.cm\epsfbox{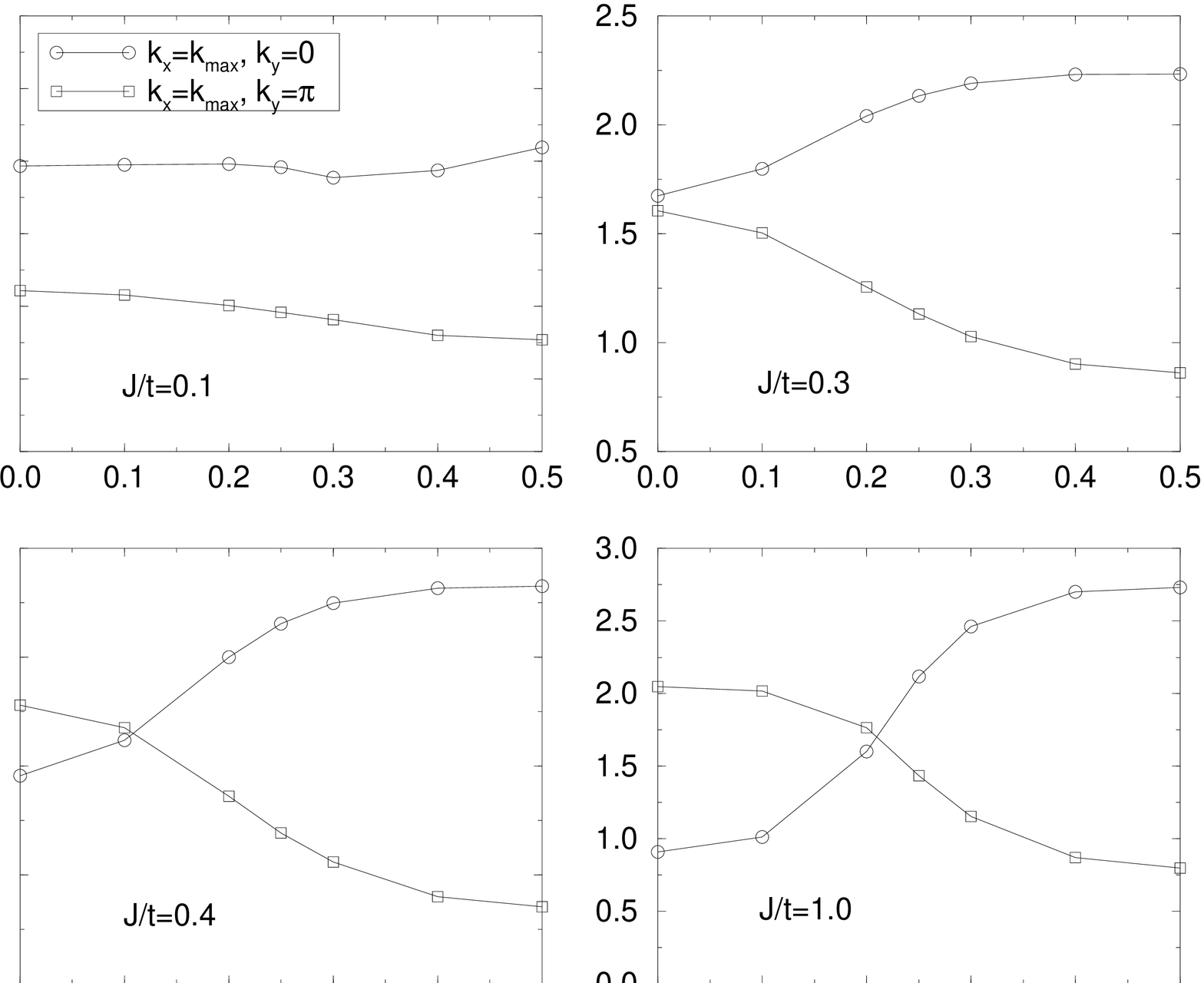}}}
\vspace{1.5cm}
\caption{The maxima of the spin--spin correlations for a 10-rung ladder
with 4 holes in Fourier space in the branches $k_y=0, \pi$ as function of
$\gamma/\pi$.} 
\label{sgmax}
\end{figure}

In Fig.~\ref{sgmax} the maxima of the spin--spin correlations as a function of
$\gamma$ are shown for the different $J/t$ values. They are at
$\bk=(\pi,0)$(circle) and \mbox{$\bk=(\frac{4\pi}{5},\pi)$}(square) 
(with the exception of
 $J/t=0.1$, where the   maximum is at \mbox{$\bk=(\frac{3\pi}{5},0)$}
when \mbox{$\gamma\ge 0.4\pi$}),
emphasizing the intraband processes in the antibonding band.
The crossing of the maxima for \mbox{$J/t\ge 0.4$} is related to the
destruction of the hole pairs. It occurs  at 
$\gamma\simeq 0.11\pi$ for \mbox{$J/t=0.4$} while for \mbox{$J/t=1.0$} the
crossover occurs 
at a higher \mbox{$\gamma \simeq 0.21\pi$} since the hole pairs are more
strongly bound.

\begin{figure}
\centerline{\rotatebox{-90}{\epsfysize=9.cm\epsfbox{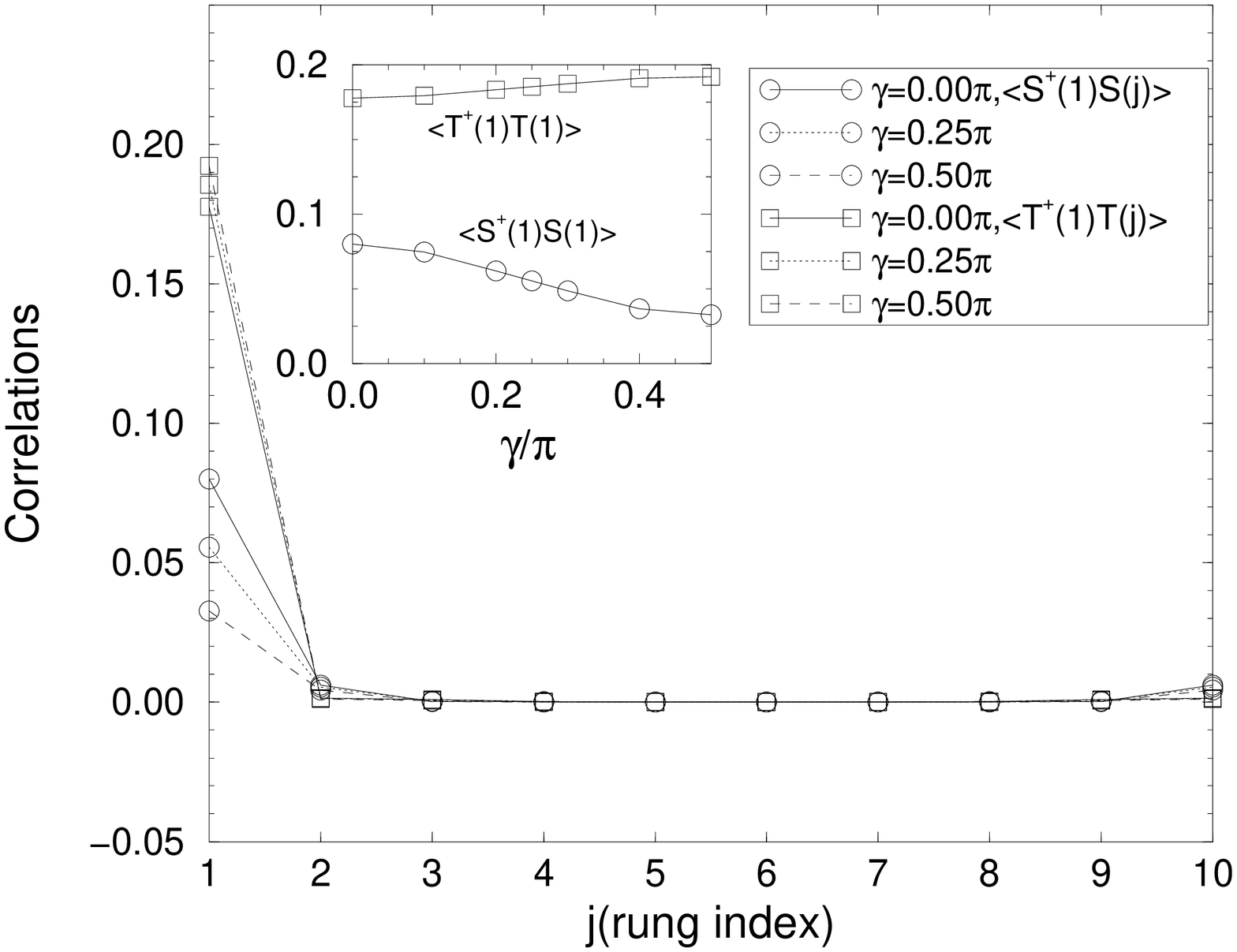}}}
\vspace{5mm}
\centerline{\rotatebox{-90}{\epsfysize=9.cm\epsfbox{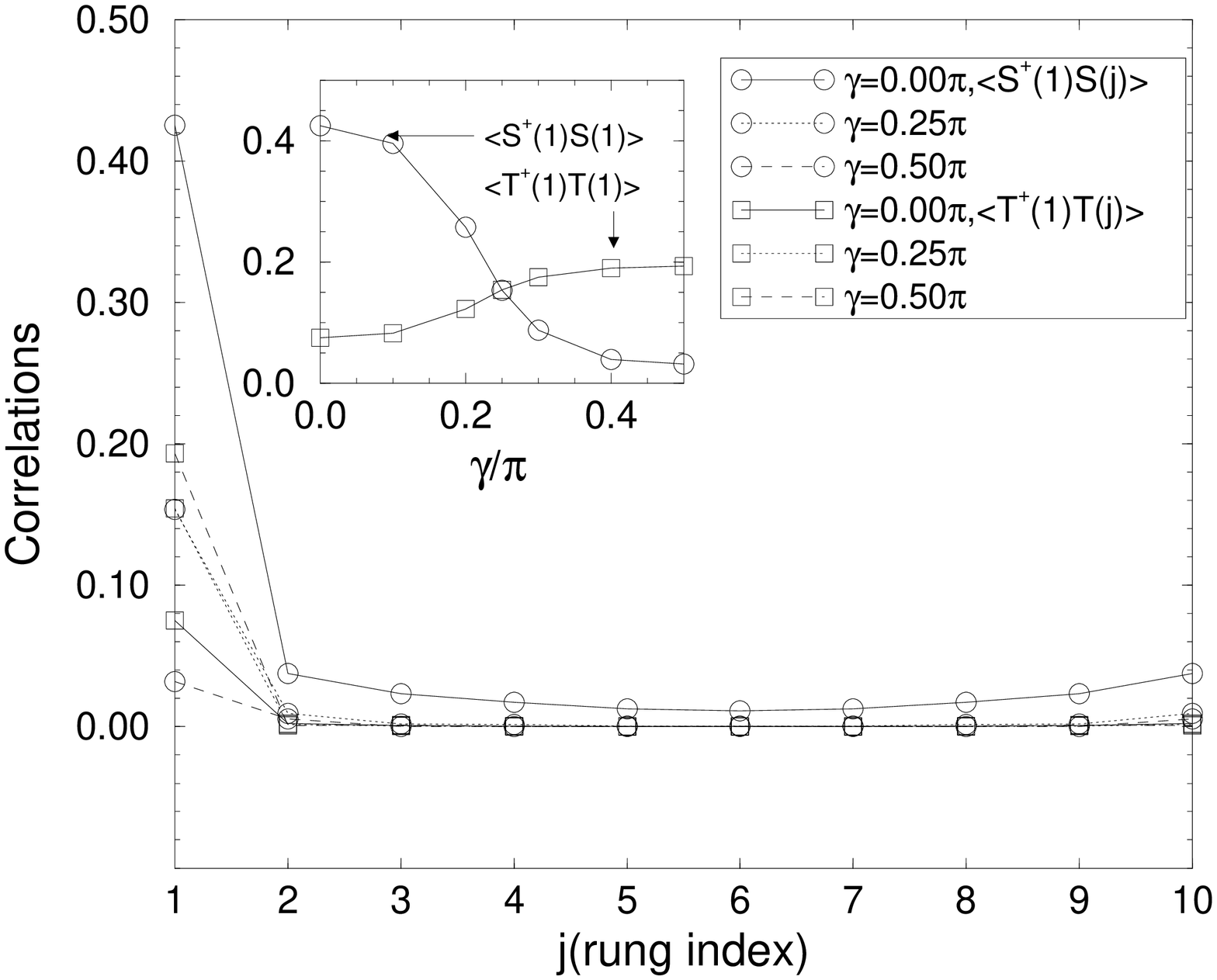}}}
\vspace{1.5cm}
\caption{The pair correlations $\SSe{1}{j}$ and $\TTo{1}{j}$ at different $\gamma$
values as function 
of the rung index for a 10-rung ladder with $n_{\mathrm{h}}=4$ holes.
The upper graph shows data for $J/t=0.1$ and the lower graph for $J/t=1.0$. 
The insets show the on-site pair correlations $\SSe{1}{1}$ and $\TTo{1}{1}$
as function of $\gamma$.} 
\label{st10h4s1s_10g}
\end{figure}

\begin{figure}
\centerline{\rotatebox{-90}{\epsfysize=9.cm\epsfbox{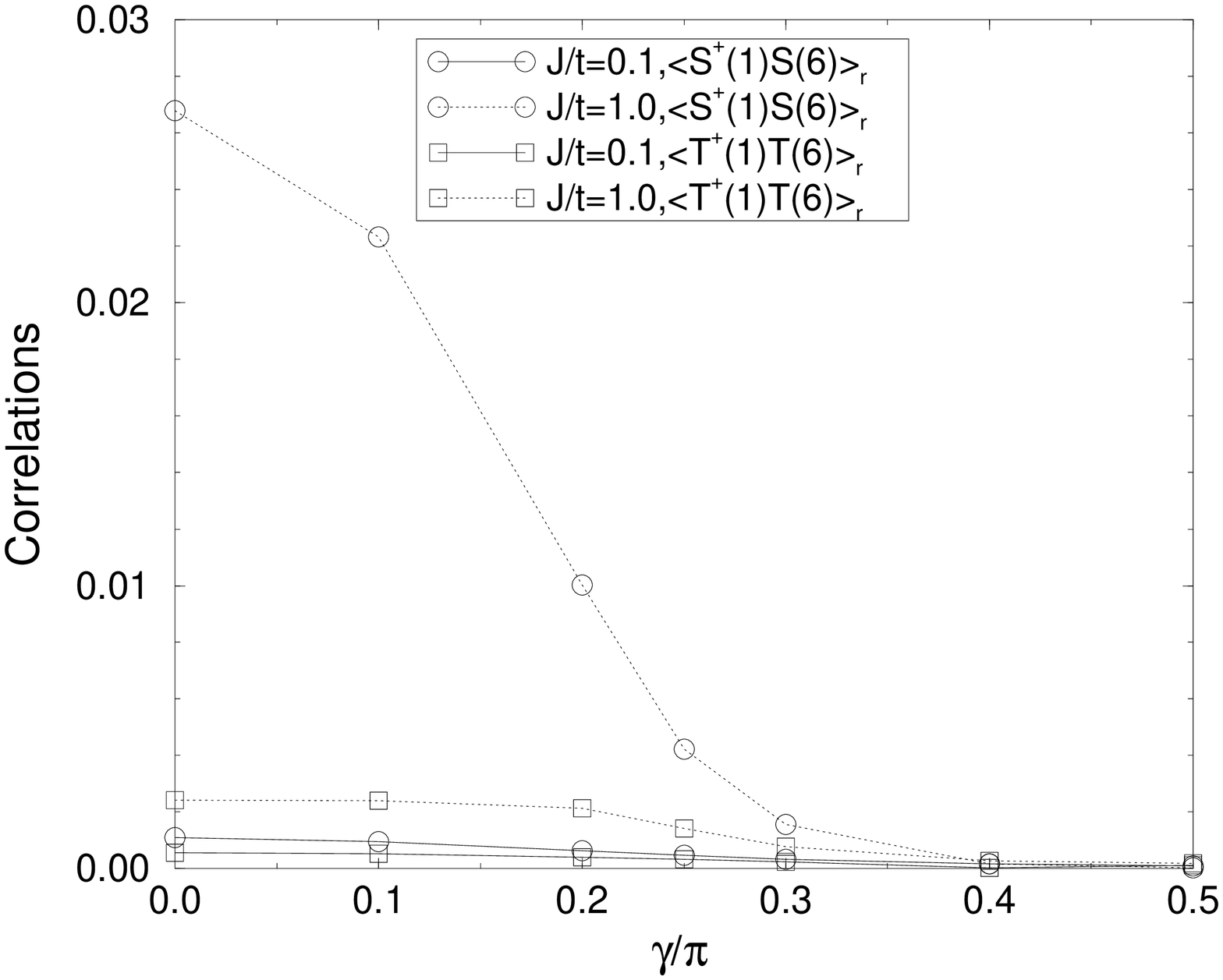}}}
\vspace{5mm}
\caption{Long-range normalized pair correlations $\SSe{1}{j}_r$ and 
$\TTo{1}{j}_r$ as
function of $\gamma$.} 
\label{st10h4s1s6g}
\end{figure}

The influence of the relative phase $\gamma$ on the singlet
and triplet pair correlation functions is  plotted in
Fig.~\ref{st10h4s1s_10g} and~\ref{st10h4s1s6g}.  In
the upper graph of Fig.~\ref{st10h4s1s_10g} the pair correlation
$\SSe{1}{j}$ and \mbox{$\TTo{1}{j}$} at \mbox{$J/t=0.1$}
are plotted for different $\gamma$'s. Pair correlations along the 
rung decrease rapidly to zero. At $j=1$ the pair correlation measures the
number of singlets (triplets) on the rung. The number of singlets is much
lower than the triplet number of the first rung, in agreement with the
destruction of the singlet liquid state. 
By switching on $\gamma$, the singlet number decreases while the
triplet number increases. The inset of the graph shows the number
operators as a function of $\gamma$. 
In the lower graph, the singlet and triplet pair correlations are measured in
the 
spin-gapped phase at $J/t=1.0$. The singlet pair correlation starts with a
higher value than the triplet pair correlation. It decreases rapidly to a
lower value and shows a finite value at $j=6$. 
When $\gamma$ is switched on, the on-site singlet is
suppressed while the triplet pair correlation is increased. The crossover
occurs at \mbox{$\gamma\simeq 0.25\pi$}. 
This on-site behavior emphasizes the picture of destroyed singlets
on the rungs. 

The long-range behavior is shown in Fig.~\ref{st10h4s1s6g}. where the
normalized  SS(TT) pair correlation at site $j=6$ are plotted as
functions of \mbox{$\gamma$}.  For \mbox{$J/t=0.1$} the pair correlations
remain nearly 
unaffected  by the introduction of $\gamma$ while for \mbox{$J/t=1.0$} the
pair correlations are suppressed and reach nearly the same value at
\mbox{$\gamma=0.5\pi$}. 

\section{Conclusions}

In this paper, an extensive study of the small  $J/t$-region at low-doping 
using exact diagonalization results of small cluster has been
performed. 
First the hole--hole correlation showed hole repulsion for small $J/t$ for
a certain doping range indicating the presence of a C2S2 phase \cite{caution}.
However, finite size effect strongly affect the result and different
correlations were introduced in order to examine other properties of the
system. 
For a very small doping ($\delta \leq 0.1$), no clear sign of the
existence of such region has been seen.
However, a very tiny region between the ferromagnetic and the LE phase
may  occur. 

\begin{figure}
\centerline{\rotatebox{-90}{\epsfysize=10.cm\epsfbox{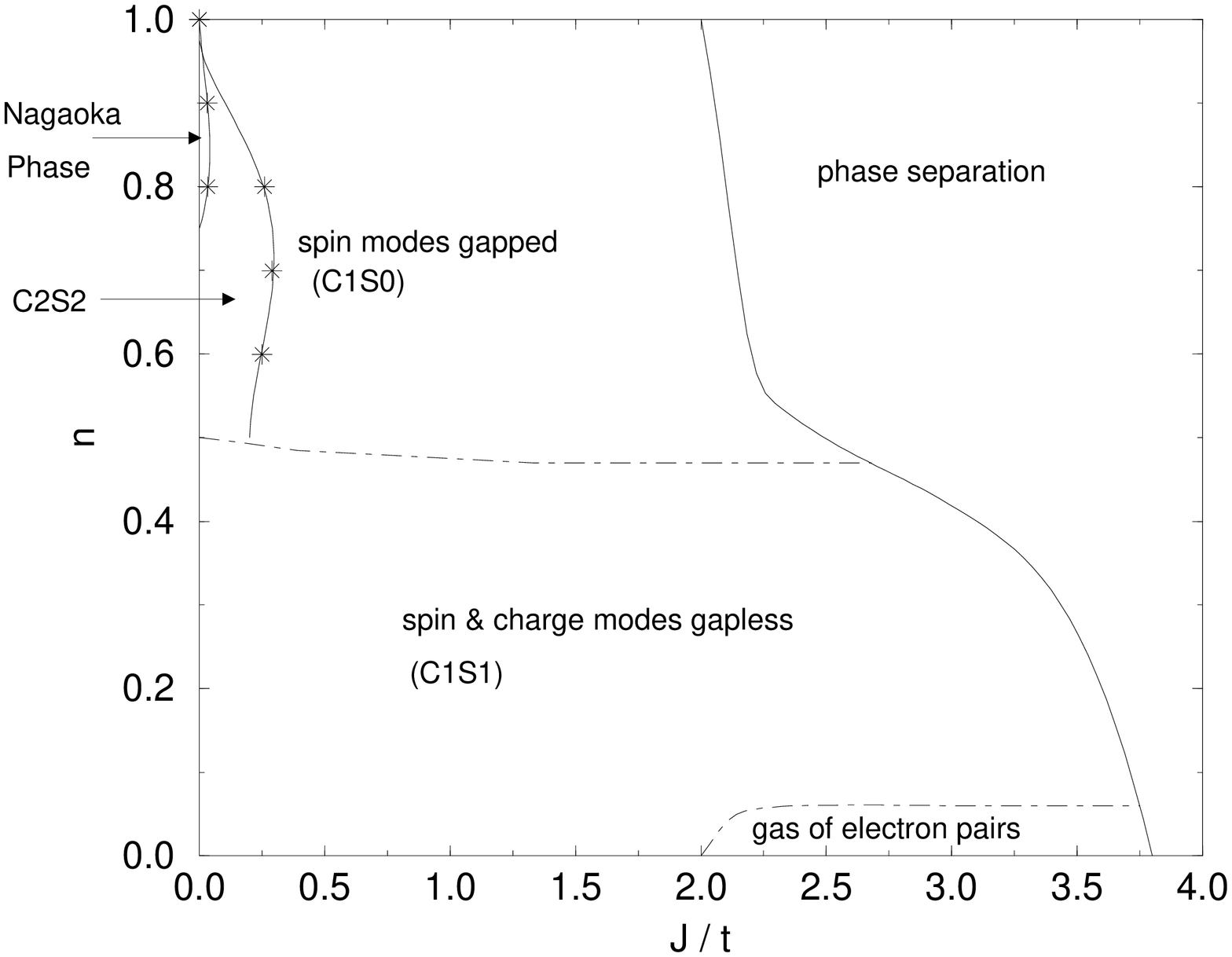}}}
\caption{The speculative phase diagram taken from the work of Poilblanc
{\it et} al. with the new phases added at low value of
$J/t$ and \mbox{$n>1/2$}. When $n<1/2$ the Fermi energy crosses only the bonding band and a 
single LL occurs.} 
\label{diaphase}
\end{figure}

For larger doping clear signs of a C2S2 phase have been observed.
Two different sets of critical values were found. 
Those obtained from the study of the crossover of the maximal value of
the spin--spin correlation function
\mbox{$\cSk(\bk)$} in Sec.~\ref{sec:sc} are consistent 
with the values from the study of the long-range singlet--singlet pair
correlation of 
Sec.~\ref{sec:ssttc}. Their mean values are
\mbox{$(J/t)_{\mathbf{c}} \simeq 0.26,0.29$}, and $0.25$
for \mbox{$\delta=0.2,0.3$}, and $0.4$, respectively.
In Fig.~\ref{diaphase}, they are included in  the previous phase diagram
proposed 
by Hayward and Poilblanc~\cite{HaywardPoil}.
In the very small $J/t$ region, the Nagaoka phase is shown.
At quarter filling, in the non-interacting limit $E_F$ just touches the
anti-bonding band while 
Umklapp processes occur in the bonding band. These features are difficult to simulate in finite 
clusters and will not be discussed here.

\section{Acknowledgments}

We wish to thank S.~Haas, C.~Hayward, D.~Poilblanc, F.~Mila for helpful
discussion. This work was supported by the ``Fond National Suisse''.
 Calculations have been performed on the IBM 
RS/6000-590 and on the DEC AXP
8400 of ETH-Z\"urich and on the NEC SX-3/24R of CSCS Manno.

\begin{appendix}
\section{Transformed Hamiltonian}
\label{appen}
The introduction of the relative phase between bonding and antibonding
operators does not transform the hopping term $H_h$ but the 
magnetic part is considerably modified. 
	The first term of the magnetic coupling along the chain
Equ.~\ref{termpar}, 
\begin{eqnarray*}
H_{\mathrm{t},\Jpar,1} & = &
	  J_{||}\sum_{i}\Bigg[ \left(1-\frac{1}{4}(1-\cos(2\gamma))\right)
\left(\bS_{1i}\bS_{1i+1} \right. \nonumber \\
	& & \left. + \bS_{2i}\bS_{2i+1}-\frac{1}{4}(n_{1i}n_{1i+1}+
n_{2i}n_{2i+1})\right)\Bigg], 
\end{eqnarray*}
is similar to the original Hamiltonian with a
renormalized coupling constant which decreases
as \mbox{$\gamma \in [0,\pi]$} increases.  For $\gamma=\pi$ the coupling
strength is  half that of $\gamma=0$. 

The second term
\begin{eqnarray*}
H_{\mathrm{t},\Jpar,2} & = &
 	 J_{||}\sum_{i}\Bigg[
\frac{1}{4}(1-\cos(2\gamma))\left(\bS_{1i}\bS_{2i+1}+ \right.
	\nonumber \\
	& & \left. \bS_{2i}\bS_{1i+1}
-\frac{1}{4}(n_{1i}n_{2i+1}+n_{2i}n_{1i+1})\right)\Bigg],
\end{eqnarray*}
is similar to the first part but with 
antiferromagnetic coupling along diagonals. 

The third part, 
\begin{eqnarray*}
	\lefteqn{H_{\mathrm{t},\Jpar,3}= J_{||}\sum_{i}\Bigg[ -\frac{1}{4}(1-\cos(2\gamma))}\nonumber \\
	& & \left[ (c^{\dag}_{1i}\bsigma c_{2i}-c^{\dag}_{2i}\bsigma c_{1i})
(c^{\dag}_{1i+1}\bsigma c_{2i+1}-c^{\dag}_{2i+1}\bsigma c_{1i+1}) \right.\\
	& & 
	\left. -\frac{1}{4}(c^{\dag}_{1i}\bun c_{2i}-c^{\dag}_{2i}\bun c_{1i})
	(c^{\dag}_{1i+1}\bun c_{2i+1}-c^{\dag}_{2i+1}\bun
	c_{1i+1})\right]\Bigg] ,\\
\end{eqnarray*}
features spin-dependent hopping terms involving
a simultaneous hopping on two neighboring rungs. 

The  last part, 
\begin{eqnarray*}
H_{\mathrm{t},\Jpar,4} & = &
	  J_{||}\sum_{i}\Bigg[-\frac{i}{4}(\sin(2\gamma)) \times \nonumber
\\ 
	& & \left[(\bS_{1i}-\bS_{2i})(c^{\dag}_{1i+1}\bsigma
c_{2i+1}-c^{\dag}_{2i+1}\bsigma c_{1i+1}) \right. \\
	& & + (c^{\dag}_{1i}\bsigma
c_{2i}-c^{\dag}_{2i}\bsigma c_{1i})(\bS_{1i+1}-\bS_{2i+1}) \nonumber \\
	& & - \frac{1}{4}(n_{1i}-n_{2i})(c^{\dag}_{1i+1}\bun
c_{2i+1}-c^{\dag}_{2i+1}\bun c_{1i+1})\\ 
	& & -\frac{1}{4}(c^{\dag}_{1i}\bun
c_{2i}-c^{\dag}_{2i}\bun c_{1i})(n_{1i+1}-n_{2i+1})\bigg] \Bigg], \\
\end{eqnarray*}
is a combination of spin and hopping operators involving a spin dependent hopping on one
rung if the neighboring rung is occupied. 

The interchain magnetic coupling after some algebra can be written as  
\begin{eqnarray*}
\lefteqn{H_{\mathrm{t},\Jperp} =  } \nonumber \\
& & J_{\perp}\sum_{i}\left(1-\frac{1}{2}(1-\cos(2\gamma))\right)\nonumber 
	\left(\bS_{1i}\bS_{2i}-\frac{1}{4}n_{1i}n_{2i}\right),
\end{eqnarray*} 
vanishing for \mbox{$\gamma=\pi$}.

\end{appendix}

\end{multicols}

\onecolumn


\begin{references}
\bibitem{bednorz}J.~G.~Bednorz and K.~A.~M\"uller, Z.\ Phys.\ B {\bf 64}, 189
(1986).

\bibitem{dagotto}E.~Dagotto, J.~Riera, and D.~Scalapino, Phys.\ Rev.\ B{\bf
 45},  5744 (1992).

\bibitem{barnes} T.~Barnes {\it et al} Phys.\ Rev.\ B {\bf 47}, 3196 (1993).



\bibitem{whiteno}S.~R.~White, R.~M.~Noack, and D.~J.~Scalapino, Phys.\ Rev.\
Lett.\ {\bf 73}, 886 (1994).

\bibitem{gopa} S.~ Gopalan, T.~M.~Rice and M.~Sigrist, Phys.\ Rev.\ B {\bf
49}, 8901 (1994).

\bibitem{Troyer}M.~Troyer, H.~Tsunetsugu, and T.~M.~Rice, Phys.\ Rev.\ B {\bf
53}, 
251 (1996);
 M.~Troyer, Simulation of constrained fermions in
low-dimensional systems, ETH-thesis No.~10793 (1994).

\bibitem{tsunetsugu} H.~Tsunetsugu, M.~Troyer, and T.~M.~Rice, Phys.\ Rev.\ 
B {\bf 51}, 16456 (1995).

\bibitem{frishmuth}B.~Frischmuth, B.~Ammon, M.~Troyer
Phys.~Rev.~B 54, R3714 (1996).


\bibitem{ricegosi} T.~M.~Rice, S.~Gopalan, and M.~Sigrist, Europhys.\ Lett.\
{\bf 23}, 445 (1994). 

\bibitem{dagottorice} E.~Dagotto, T.~M.~Rice, Science, {\bf 271}, 618 (1996).

\bibitem{carron} E.~M.~McCarron {\it et al.}, Mat.~Res.~Bull.~, {\bf 23}, 1355
(1988). 

\bibitem{siegrist}T.~Siegrist  {\it et al.}, Mat.~Res.~Bull.~, {\bf 23}, 1429 (1988). 

\bibitem{osafune}  T.~Osafune, N.~Motoyama,  H.~Eisaki and S.~Ushida, 
Phys.\ Rev.\ Lett., {\bf 78}, 1980 (1997).


\bibitem{mizuno} Y.~Mizuno, T.~Tohyama, and S.~Maekawa,  
J.\ Phys.\ Soc.\ Japan, {\bf 4}, 937 (1997).


\bibitem{magishi} K.~Magishi {\it et al} to appear in Phys.\ Rev.\ B
{\bf }.

\bibitem{uehara}M.~Uehara {\it et al.}, J. of. Phys. Jpn, {\bf 65},2764 
(1997).

\bibitem{Motoyama}For a study of the resistivity in the normal state, 
see N.~Motoyama {\it et al}, Phys.\ Rev.\ B {\bf 55},R 3386 (1997).

\bibitem{BalentsFisher}L.~Balents and M.~P.~A.~Fisher, Phys.\ Rev.\ B, 
{\bf 53}, 12133 (1996).

\bibitem{Noacketal} R.~M.~Noack, S.~R.~White, and D.~J.~Scalapino, 
Physica C, {\bf 270}, 281 (1996).

\bibitem{anderson} P.~W.~Anderson, Science {\bf 235}, 1196 (1987).


\bibitem{ZhangRice}F.~C.~Zhang and T.~M.~Rice, Phys.\ Rev.\ B {\bf 37},
3759 (1988).


\bibitem{WhiteScal}S.~R.~White and D.~J.~Scalapino, Phys.\ Rev.\ B {\bf 55},
6504 (1997).

\bibitem{PoilblancScal}D.~Poilblanc, D.~J.~Scalapino and W.~Hanke,
Phys.\ Rev.\ B {\bf 52}, 6796 (1995).

\bibitem{HaywardPoil}C.~A.~Hayward and D.~Poilblanc, Phys.\ Rev.\ B {\bf 53},
1 (1996).

\bibitem{lowestBC} They  
correspond to boundary conditions with a closed shell but do not
necessarily correspond to the BC yielding the lowest energy in the
non-interacting limit. 

\bibitem{Nagaoka}Y.~Nagaoka, Phys.\ Rev.\ {\bf 147}, 392 (1965), 
Y.~Nagaoka, Solid Stat.\ Com.\ \ {\bf 3}, 409 (1965). 


\bibitem{specialcare} Special care has to be taken for the
\mbox{$n_{\mathrm{h}}=8$} case. Both  
APBC and PBC correspond to CSBC. The lowest ground state energy occur
with APBC. However, APBC  does {\it not} occupy states in the
antibonding band. This is a peculiar feature for this filling coming from 
the discrete set of available states in Fourier space. $E_F$
is  
close to the first available state of the antibonding band and the interband 
scattering may be enhanced. However, this case is not related to the present
study. In order to constrain the system to have at least one antibonding state
filled 
(in the non-interacting limit), PBC will be preferred.

\bibitem{phasenote} The reader should note that these pair correlations involve correlators
which are  not in  Hermitian form.  Actually, when a phase shift
is introduced in the system either through a non-zero momentum or through 
twisted boundary conditions, pair correlations will measure the corresponding
phase 
shift due to the translation from site $j$ to $1$ and will become complex. 
	No new information is contained in this phase factor thus  the
absolute value of the pair correlations will be plotted. 

\bibitem{Haldane}F.~D.~M.~Haldane in Proceedings of the International School of Physics
``Enrico Fermi'', Course CXXI, Varenna Summer School 1992. 
North-Holland Elsevier Science Publishers B.~V., Amsterdam, 1994. 

\bibitem{caution}Note  these  results are for finite clusters and thus 
that strictly speaking they show only the existence of a region 
with a very small gap. 
\end{references}
\end{document}